\documentclass[twocolumn]{aastex63}
\usepackage{appendix}
\usepackage{natbib}
\usepackage{subfigure}
\usepackage{graphicx}
\shorttitle{AGN Among Extreme UV Emitters}
\usepackage[encapsulated]{CJK}

\usepackage{amsmath} 

\newcommand{\parentnum}{205}
\newcommand{\emitternum}{12}

\usepackage[encapsulated]{CJK}

\begin{document}

\title{UNCOVERing the High-redshift AGN Population among Extreme UV Line Emitters}
\correspondingauthor{Helena Treiber}
\email{lena.treiber@princeton.edu}

\author[0000-0003-0660-9776]{Helena Treiber}
\altaffiliation{NSF Graduate Research Fellow}
\affiliation{Department of Astrophysical Sciences, Princeton University, 4 Ivy Lane, Princeton, NJ 08544, USA}

\author[0000-0002-5612-3427]{Jenny E. Greene}
\affiliation{Department of Astrophysical Sciences, Princeton University, 4 Ivy Lane, Princeton, NJ 08544, USA}

\author[0000-0003-1614-196X]{John R. Weaver}
\affiliation{Department of Astronomy, University of Massachusetts, Amherst, MA 01003, USA}

\author[0000-0001-8367-6265]{Tim B. Miller}
\affiliation{Center for Interdisciplinary Exploration and Research in Astrophysics (CIERA) and Department of Physics \& Astronomy, Northwestern University, IL 60201, USA}

\author[0000-0001-6278-032X]{Lukas J. Furtak}
\affiliation{Department of Physics, Ben-Gurion University of the Negev, P.O. Box 653, Beer-Sheva 8410501, Israel}

\author[0000-0003-4075-7393]{David J. Setton}\thanks{Brinson Prize Fellow}
\affiliation{Department of Astrophysical Sciences, Princeton University, 4 Ivy Lane, Princeton, NJ 08544, USA}

\author[0000-0001-9269-5046]{Bingjie Wang (\begin{CJK*}{UTF8}{gbsn}王冰洁\ignorespacesafterend\end{CJK*})}
\affiliation{Department of Astronomy \& Astrophysics, The Pennsylvania State University, University Park, PA 16802, USA}
\affiliation{Institute for Computational \& Data Sciences, The Pennsylvania State University, University Park, PA 16802, USA}
\affiliation{Institute for Gravitation and the Cosmos, The Pennsylvania State University, University Park, PA 16802, USA}

\author[0000-0002-2380-9801]{Anna de Graaff}
\affiliation{Max-Planck-Institut f\"ur Astronomie, K\"onigstuhl 17, D-69117, Heidelberg, Germany}

\author[0000-0001-5063-8254]{Rachel Bezanson}
\affiliation{Department of Physics \& Astronomy and PITT PACC, University of Pittsburgh, Pittsburgh, PA 15260, USA}

\author[0000-0003-2680-005X]{Gabriel Brammer}
\affiliation{Cosmic Dawn Center (DAWN), Niels Bohr Institute, University of Copenhagen, Jagtvej 128, København N, DK-2200, Denmark}

\author[0000-0002-7031-2865]{Sam E. Cutler}
\affiliation{Department of Astronomy, University of Massachusetts, Amherst, MA 01003, USA}

\author[0000-0001-8460-1564]{Pratika Dayal}
\affiliation{Kapteyn Astronomical Institute, University of Groningen, P.O. Box 800, 9700 AV Groningen, The Netherlands}

\author[0000-0002-1109-1919]{Robert Feldmann}
\affiliation{Department of Astrophysics, University of Zurich, Winterthurerstrasse 190, CH-8057 Zurich, Switzerland}

\author[0000-0001-7201-5066]{Seiji Fujimoto}\altaffiliation{NHFP Hubble Fellow}
\affiliation{
Department of Astronomy, The University of Texas at Austin, Austin, TX 78712, USA
}

\author[0000-0003-4700-663X]{Andy D. Goulding}
\affiliation{Department of Astrophysical Sciences, Princeton University, 4 Ivy Lane, Princeton, NJ 08544, USA}

\author[0000-0002-5588-9156]{Vasily Kokorev}
\affiliation{Kapteyn Astronomical Institute, University of Groningen, P.O. Box 800, 9700 AV Groningen, The Netherlands}

\author[0000-0002-2057-5376]{Ivo Labbe}
\affiliation{Centre for Astrophysics and Supercomputing, Swinburne University of Technology, Melbourne, VIC 3122, Australia}

\author[0000-0001-6755-1315]{Joel Leja}
\affiliation{Department of Astronomy \& Astrophysics, The Pennsylvania State University, University Park, PA 16802, USA}
\affiliation{Institute for Computational \& Data Sciences, The Pennsylvania State University, University Park, PA 16802, USA}
\affiliation{Institute for Gravitation and the Cosmos, The Pennsylvania State University, University Park, PA 16802, USA}

\author[0000-0001-9002-3502]{Danilo Marchesini} \affiliation{Department of Physics \& Astronomy, Tufts University, MA 02155, USA}

\author[0000-0003-2804-0648]{Themiya Nanayakkara}
\affiliation{Centre for Astrophysics and Supercomputing, Swinburne University of Technology, PO Box 218, Hawthorn, VIC 3122, Australia}

\author[0000-0002-7524-374X]{Erica Nelson}
\affiliation{Department for Astrophysical and Planetary Science, University of Colorado, Boulder, CO 80309, USA}

\author[0000-0002-9651-5716]{Richard Pan}
\affiliation{Department of Physics \& Astronomy, Tufts University, 574 Boston Avenue, Medford, MA 02155, USA}

\author[0000-0002-0108-4176]{Sedona H. Price}
\affiliation{Department of Physics \& Astronomy and PITT PACC, University of Pittsburgh, Pittsburgh, PA 15260, USA}

\author[0000-0002-9337-0902]{Jared Siegel}
\altaffiliation{NSF Graduate Research Fellow}
\affiliation{Department of Astrophysical Sciences, Princeton University, 4 Ivy Lane, Princeton, NJ 08544, USA}

\author[0000-0002-1714-1905]{Katherine A. Suess}
\altaffiliation{NHFP Hubble Fellow}
\affiliation{Kavli Institute for Particle Astrophysics and Cosmology and Department of Physics, Stanford University, Stanford, CA 94305, USA}

\author[0000-0002-9651-5716]{Katherine E. Whitaker}
\affiliation{Department of Astronomy, University of Massachusetts, Amherst, MA 01003, USA}
\affiliation{Cosmic Dawn Center (DAWN), Denmark}

\begin{abstract}
JWST has revealed diverse new populations of high-redshift ($z\sim4-11$) AGN and extreme star-forming galaxies that challenge current photoionization models. In this paper, we use rest-frame UV emission-line diagnostics to identify AGN candidates and other exceptional ionizing sources, complementing previous studies predominantly focused on broad-line AGN. From a parent sample of \parentnum~$\mathrm{z_{spec}}>3$ UNCOVER galaxies with NIRSpec/PRISM follow-up, we identify 12 galaxies with C IV, He II, and/or C III] emission. Three of these galaxies also exhibit clear N III] and/or N IV] lines.
Leveraging the combined rest-optical and UV coverage of PRISM, we limit the emission-line model space using the sample’s [O III]/H$\beta$ distribution, significantly decreasing the overlap between AGN and star-formation models in the UV diagnostics. We then find that the five He II emitters are the strongest AGN candidates, with further support from two [Ne V] detections and one X-ray detection from \emph{Chandra}. 
Our Balmer line fits also reveal one new broad-line AGN at z=6.87.
We cannot robustly quantify the AGN fraction in this sample, but we note that close to 20\% of $\mathrm{M_{*}>2\times10^{9}\,M_{\odot}}$ parent sample galaxies are AGN candidates. The lower-mass line emitters, which are consistent with both AGN and star-forming photoionization models, have more compact sizes and higher specific star formation rates than the parent sample. Higher-resolution and deeper data on these UV line emitters should provide much stronger constraints on the obscured AGN fraction at $z > 3$.
\end{abstract}

\section{Introduction}
Supermassive black holes (SMBHs) are ubiquitous in massive galaxies \citep{Kormendy1995,Richstone1998}. 
Actively accreting SMBHs (i.e., active galactic nuclei, or AGN) provide an opportunity for detailed studies of accretion and shape our understanding of the co-evolution of SMBHs and their host galaxies \citep[][]{Ho2008,Haggard2010,Heckman2014,Lacerda2020}.

In the study of this co-evolution, high-redshift AGN are crucial, placing constraints on SMBH seeding \citep[e.g.,][]{Volonteri2016,Dayal2019,Greene2020,Inayoshi2020,Bhowmick2022,Li2023,Zhang2023} as well as AGN feedback and their contribution to reionization \citep[e.g.,][]{Fan2006,Madau2015,Trebitsch2023,Dayal2024,Madau2024}.  
However, the identification of high-redshift AGN is challenging, partially because low-metallicity gas and stellar populations decrease the efficacy of traditional AGN line ratio diagnostics \cite[e.g.,][]{Shapley2005,Groves2006,Kewley2013o,Kewley2013t,Juneau2014,Steidel2014,Feltre2016,Nakajima2022,Hirschmann2019,Hirschmann2023}.

Before the James Webb Space Telescope \citep[JWST,][]{Gardner2023}, the most fruitful approaches to search for high-redshift AGN were UV, X-ray, and radio selection, each with their own biases \citep[for a review, see][]{Fan2023}. From these searches, we knew of a couple hundred AGN at $z\gtrsim6$ \citep[e.g.,][]{Banados2016,Wang2017}. 
However, the handful of pre-JWST $z\gtrsim7$ AGN all have broad lines \citep{Mortlock2011,Banados2018,Fan2019,Wang2021} and spectroscopic follow-up revealed rest-frame UV features comparable to those of the \cite{VandenBerk2001} lower-redshift broad-line SDSS composite. We are therefore missing a census of the critical population of high-redshift narrow-line AGN. For these objects, which likely comprise the majority of the overall population, there is obscuration of the region of SMBH gravitational influence (i.e., the broad-line region), preventing observations of the AGN disk continuum and the $>1000 \, \mathrm{km \, s^{-1}}$ gas velocity widths \citep[e.g.,][]{Hickox2018}.

Selections of objects with strong rest-frame UV lines have yielded narrow-line AGN at cosmic noon (z$\sim$2$-$3), where the Lyman break and rest-frame UV are observable with ground-based optical telescopes \citep[e.g.,][]{Hainline2012,Alexandroff2013}. These observations highlight the different UV features typical of the broad- and narrow-line regions, which extend to $\lesssim$1 pc and from $\sim$100 pc to $\sim$1 kpc, respectively \citep{Hickox2018}.
For successful selection of narrow-line AGN beyond cosmic noon, further work is necessary for the identification of effective diagnostics. 
Plaguing higher-redshift diagnostics are the uncertainties and degeneracies arising from the potential contribution of Wolf-Rayet stars, supernova-driven winds, shocks, and Population III stars \citep{Hirschmann2019}. 
As a result, studies of lower-redshift metal-poor galaxies are essential, as they serve as analogues for higher redshift star-forming contaminants with stellar ionization signatures that we could mistake for AGN \citep{Erb2010,Stark2014,Mingozzi2023}. 
These empirical approaches are complementary to theoretical efforts to separate AGN and star-forming photoionization models in parameter space, regardless of metallicity \citep[e.g.,][]{Feltre2016,Gutkin2016,Hirschmann2019,Calabro2023}.

JWST is central to the efforts to identify more high-redshift AGN and constrain the trustworthiness of potential selection methods. Thanks to its wavelength range and unprecedented sensitivity, this facility has already both pushed the AGN redshift record above 10 \citep{Bogdan2023,Goulding2023,Maiolino2024GNz11} and provided a more representative sample of $z\gtrsim4$ AGN. 
These AGN have been selected for their broad lines \citep{Harikane2023,Kocevski2023,Larson2023,Maiolino2023,Ubler2023}, color and morphology \citep{Endsley2022,Onoue2023,Ono2023,Furtak2023triple,Yang2023}, and X-ray luminosity \citep{Bogdan2023,Goulding2023,Kocevski2023Xray}.

Unlike the previous record-holders, many JWST AGN are undetected in X-rays, UV-faint, and display significant reddening \citep[e.g.,][]{Yue2024,Maiolino2024}.
One common new population of AGN is the subset of so-called ``Little Red Dots" (LRDs) confirmed to have broad lines \citep{Furtak24Nature,Greene2024,Matthee2024,Wang2024,Lin2024}. LRDs are generally characterized by compactness and a continuum that is blue in the rest-frame UV and red in the rest-frame optical \citep{Labbe2023,Barro2024,Akins2024,Kocevski2024,Kokorev2024}.

Despite the successes of these selections, we know we are missing a critical population of narrow-line AGN.
To this end, in conjunction with photoionization models, \cite{Scholtz2023} and \cite{Mazzolari2024} identified a total of 59 $z>3$ narrow-line AGN candidates using a combination of rest-frame UV and optical lines.
In this paper, we apply the \cite{Feltre2016} UV diagnostics to UNCOVER \citep{Bezanson2022} galaxies to better understand the narrow-line and UV-selected AGN populations at high-redshift. 
We highlight the power of the NIRSpec/PRISM data to identify and characterize line emitters, even when the low resolution blends many relevant lines.
In particular, we leverage the rest-frame optical coverage to significantly cut down on the relevant star-forming models in the UV diagrams. We also use simulated PRISM spectra to demonstrate our ability to 
select high-EW AGN candidates.

The paper is organized as follows. In Section \ref{sec:uncover}, we describe the JWST UNCOVER program, the PRISM reduction pipeline, and our parent sample of UNCOVER $z>3$ sources. We outline our line fitting procedures in Section \ref{sec:fitting}.
In Section \ref{sec:diagram_intro}, we present the UV diagnostic diagrams, their associated photoionization models, and the locations of literature star-forming galaxies, AGN, and ambiguous cases like GN-z11.
In Section \ref{sec:diagnostics}, we then place our sample of line emitters on these and other diagnostics to identify the subset of AGN candidates. The line emitters also have noteworthy host galaxy properties, as we explore in Section \ref{sec:properties}.
In Section \ref{sec:discussion}, we discuss the possible origins of the detected lines and conclude with an outlook on further possibilities for identifying and separating high-redshift narrow-line AGN from other sources of hard ionizing spectra.

\section{UNCOVER Imaging and Spectroscopic Follow-Up}
\label{sec:uncover}
In this section, we provide an overview of UNCOVER \citep{Bezanson2022} and the reduction of the PRISM spectroscopic follow-up used in this paper. 

\subsection{UNCOVER Overview}
\begin{figure}
    \centering
    \includegraphics[width=\columnwidth]{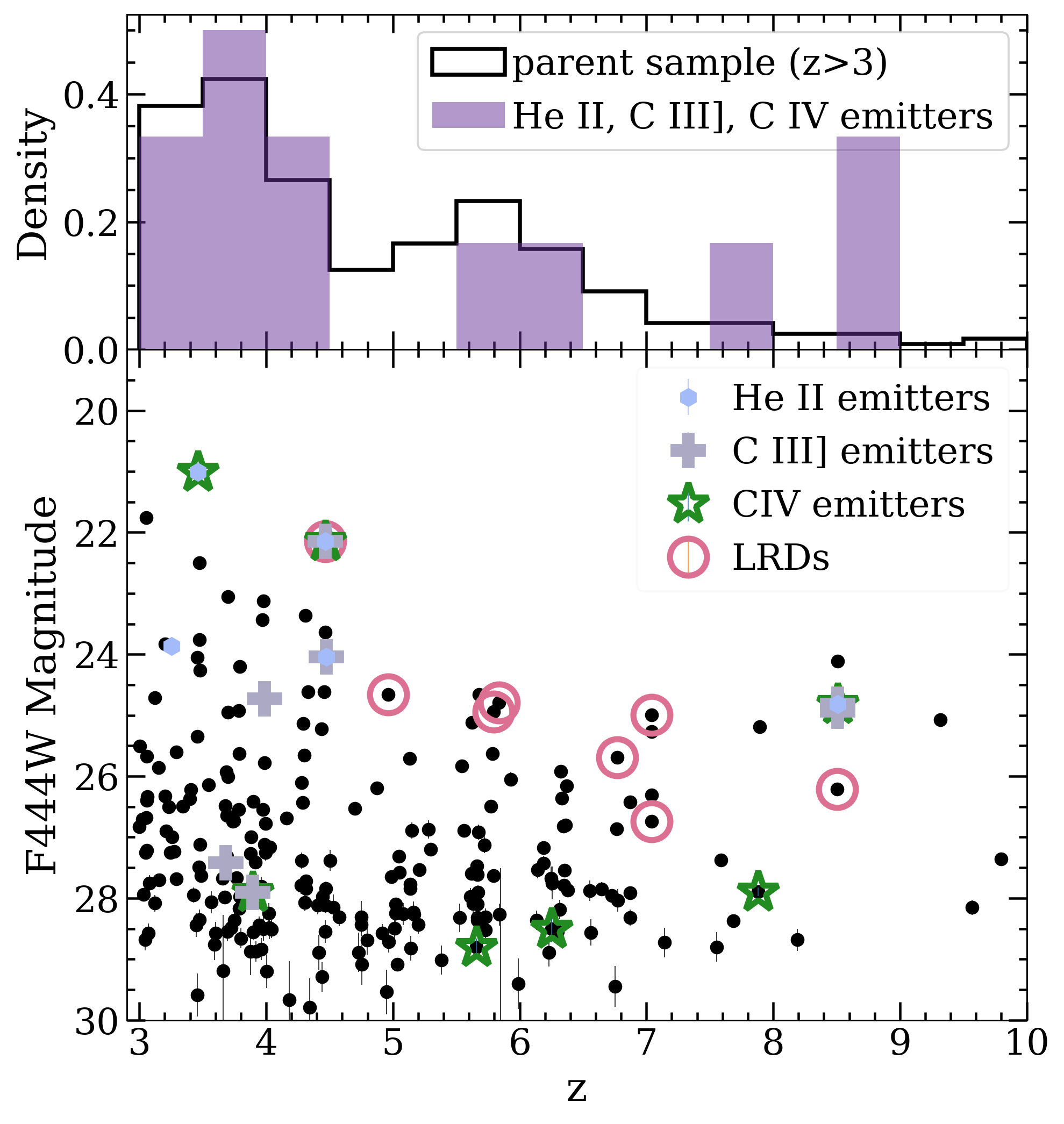}
    \caption{Apparent magnitude in the NIRCam F444W filter vs. spectroscopic redshift for the parent sample of $z>3$ UNCOVER galaxies. We mark the locations of \cite{Greene2024} LRDs and our line-emitting sample. In the top panel, we show the redshift distributions for the parent sample and for the sample of emitters.}
    \label{fig:mag_redshift}
\end{figure}

\setlength{\tabcolsep}{4pt}

\begin{table*}
\centering
\begin{tabular}{cccccccccc}
\hline
MSA ID & ra [deg] & dec [deg] & $z_{\mathrm{spec}}$ & $\mu$ & F444W [mag] & log($\mathrm{M_{*}/M_{\odot}}$) & log(sSFR/$\mathrm{yr^{-1}}$) & $\mathrm{r_{eff}}$ [arcsec] & log([O III]/H$\beta$) \\
\hline
\hline
45209&3.555938&$-30.344751$&3.001&1.68&26.83$\pm$0.06&$7.7^{+0.5}_{-0.2}$&$-8.3^{+0.2}_{-0.4}$&$0.04^{+0.01}_{-0.01}$&$0.97^{+0.11}_{-0.09}$\\
36678&3.594799&$-30.360729$&3.004&1.83&25.51$\pm$0.02&$8.5^{+0.1}_{-0.1}$&$-8.4^{+0.2}_{-0.2}$&$0.13^{+0.01}_{-0.01}$&$0.98^{+0.14}_{-0.12}$\\
46973&3.572151&$-30.341526$&3.025&1.75&26.71$\pm$0.06&$8.2^{+0.2}_{-0.2}$&$-8.7^{+0.3}_{-0.2}$&\dots&$0.83^{+0.09}_{-0.08}$\\
11126&3.610193&$-30.405520$&3.047&2.09&28.68$\pm$0.18&$7.1^{+0.4}_{-0.3}$&$-8.5^{+0.3}_{-0.4}$&$0.40^{+0.25}_{-0.15}$&$>0.35$\\
\dots&\dots&\dots&\dots&\dots&\dots&\dots&\dots&\dots&\dots\\
\hline
\end{tabular}
\caption{Properties of the parent sample. Stellar masses are corrected for magnification using the given $\mu$. Specific star formation rate (sSFR) is averaged over 100 Myr. We measure effective radius in the F444W band. Only a section of the table
is shown here; the rest can be found in the ancillary files.}
\label{tab:parent}
\end{table*}

UNCOVER is a JWST Cycle 1 Treasury program consisting of ultradeep NIRCam imaging and spectroscopic follow-up of $\sim$600 sources \citep{Price2024}.
The Cycle 1 imaging covers 45 arcmin$^2$ in the CLASH/Hubble Frontier Fields z=0.308 cluster Abell 2744 \citep{Lotz2017}, where the boost from strong gravitational lensing allows for an effective depth of $\sim$31.5 rather than $\sim$30 mag. 
Much of the field also has medium band imaging from Cycle 2 \citep{Suess2024}.
Both the lens model \citep{Furtak2023,Price2024} and the DR3 images and catalogs are publicly available \citep{Weaver2023,Suess2024}. \cite{Wang2024sps} presents the \texttt{prospector} \citep{Leja2017,Johnson2021,Wang2023pbeta} photometric redshifts and estimates of the physical properties.

Complementary to the JWST observations are multi-wavelength imaging campaigns of the UNCOVER field, including deep radio/sub-mm and X-ray observations that are critical to identifying AGN. ALMA 1.2 mm continuum photometry was extracted for all UNCOVER sources, with continuum RMS sensitivity $\lesssim$33$\mu$Jy \citep[DUALZ,][]{Fujimoto2023}. 
The cluster also has deep ($\sim$2.3 Ms) \emph{Chandra} ACIS-I imaging (PI:A.Bogdán), which pushed the AGN redshift frontier with the discovery of a z = 10.1 AGN \citep{Bogdan2023,Goulding2023}. 

Phase two of the UNCOVER program includes NIRSpec/PRISM follow-up of $\sim$600 sources of interest. With key science and legacy goals in mind, the extensive ranking of follow-up priorities included, for example, the highest-redshift candidates, $z>4$ dusty or quiescent galaxy candidates, and $z>6$ AGN candidates \citep{Labbe2021}. However, the majority of follow-up spectroscopic targets simply sample galaxies across a range of stellar mass and redshift. \cite{Price2024} detail the UNCOVER spectroscopic targeting strategy and spectroscopic data reduction.

\subsection{NIRSpec/PRISM Data Reduction}
Data reduction of the PRISM spectroscopy with \texttt{msaexp} \citep[v0.6.10][]{Brammer2022} includes the following steps \citep{Price2024}. Starting with the MAST1 level2 products, msaexp corrects 1/f noise, masks snowballs and other artifacts, and subtracts bias from individual exposures. Next, the pipeline applies WCS, identifies the slits, flat-fields, and performs path-loss correction. 
On a common grid, 2D slits are extracted and drizzled. We then vertically shift and stack the 2D spectra before applying a local background subtraction. This extraction uses a Gaussian model with a free mean and width \citep[][]{Horne1986}. 

In order to account for the magnification of background galaxies due to strong gravitational lensing by the Abell~2744 galaxy cluster, we use the UNCOVER \texttt{v2.0} strong lensing model \citep{Furtak2023}, which was recently updated with 14 additional multiple image redshifts and 152 new cluster members \citep{Price2024}. The model was constructed with an updated version of the \citet{Zitrin2015} analytic code and includes five smooth cluster-scale dark matter halos and 552 cluster member galaxies. We refer the reader to \citet{Furtak2023} for details on the model parametrization and optimization. The \texttt{v2.0} model achieves an average lens plane image reproduction error of $\Delta_{\mathrm{RMS}}=0.60\arcsec$.

\subsection{Parent Sample}
The NIRSpec/PRISM spectra provide a wide range of emission lines ($\mathrm{\lambda_{obs} \, 0.6 - 5.3 \, \mu m}$) at low resolution (R$\sim$50$-$400). As such, for $4 < z < 10$, the spectra include, at minimum, Ly$\alpha$ through H$\beta$. Because we require C IV for the \cite{Feltre2016} diagnostics (Section \ref{sec:diagnostics}), we start with only the UNCOVER spectroscopic follow-up sources with $\mathrm{z_{spec} > 3}$ (and redshift quality flag of 2 or 3 in \cite{Price2024}), leaving a parent sample of \parentnum~galaxies. In Figure \ref{fig:mag_redshift}, we provide the distribution of magnitudes as a function of redshift for the parent sample and highlight the ``Little Red Dots" (LRDs) from \cite{Greene2024} in addition to the line emitters that we present in Section \ref{sec:diagnostics}. 
Table \ref{tab:parent} includes key properties of the parent sample.

To obtain masses and star formation rates, we use the \texttt{prospector} \citep{Leja2017,Johnson2021} fits to the NIRCam photometry, as outlined in \cite{Wang2024sps}. 
Because of the complex spectroscopic selection and the resulting bias of the parent sample, we also compare the masses and star formation rates to the 16,665 UNCOVER galaxies with $\mathrm{z_{phot}>3}$. 
Wherever available, we incorporate the medium band photometry from the Cycle 2 program ``Medium Bands, Mega Science" \citep{Suess2024}.

To measure the effective radius of each galaxy, we use \texttt{pysersic} \citep{Pasha2023} to fit a single Sérsic profile to the F444W image, taken with a pixel scale of 0.04 arcseconds per pixel. We calculate uncertainties assuming the parameter posteriors are Gaussian, using the Laplace Approximation. 
After converting the radii in pixels to arcseconds, we use \cite{Planck2020} cosmological parameters to obtain radii in kiloparsecs.
We correct all these quantities with the magnifications.

\section{Line Fitting}
\label{sec:fitting}
\begin{figure*}
    \centering
    \includegraphics[width=2\columnwidth]{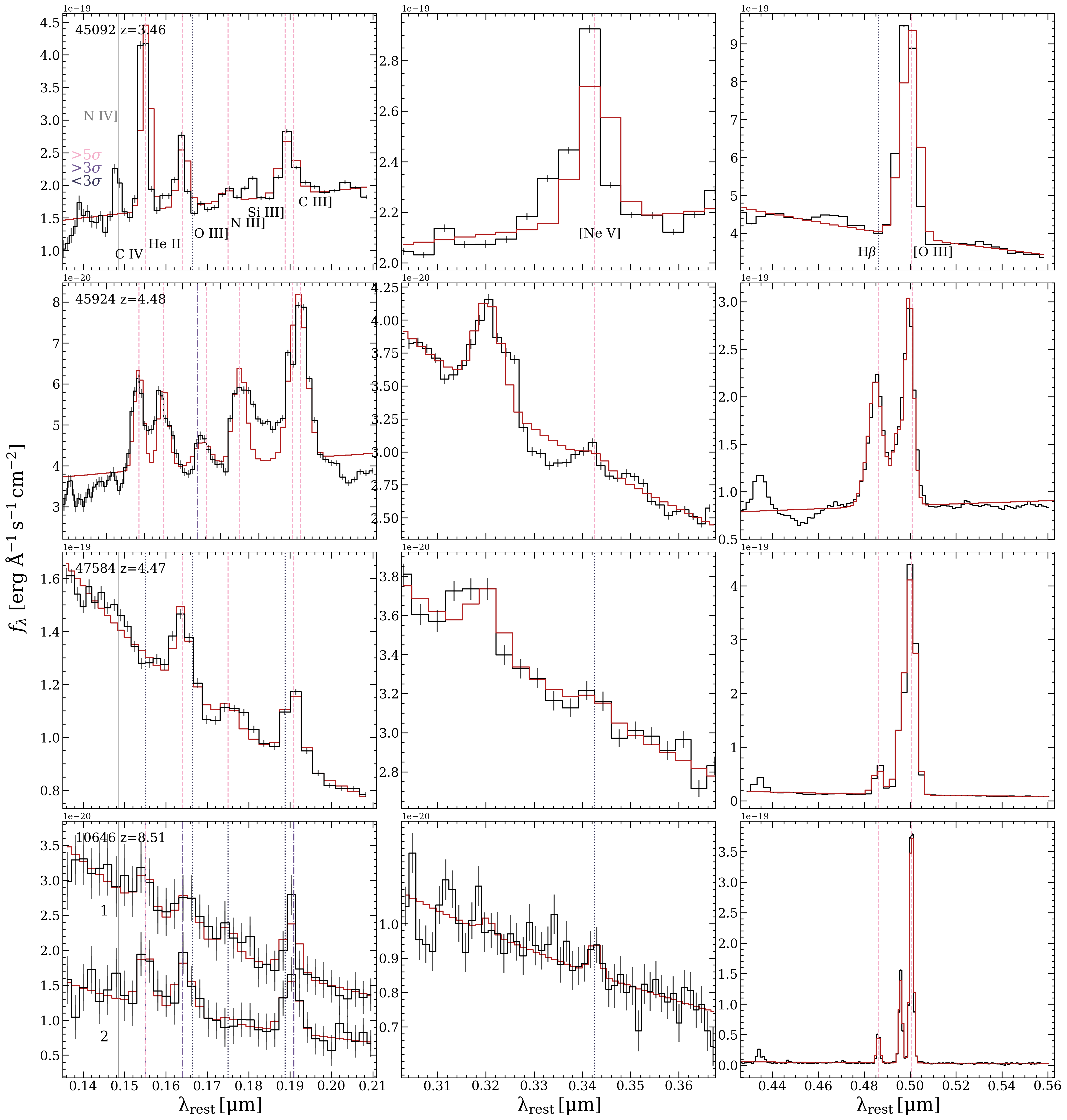}
    \caption{From left to right, we show rest-frame UV, [Ne V], and H$\beta$+[O III] data (black) and fits (red) for the best sources in the line-emitting sample. We mark fitted lines with vertical lines, highlighting the He II+O III] and Si III]+C III] blends. Pink dashed lines, purple dash-dot lines, and navy dotted lines indicate $>5\sigma$, $3-5\sigma$, and $<3\sigma$ line fits, respectively. 
    The continuum near [Ne V] is not well-described by a power law, but the two $\gtrsim 5\sigma$ detections (45092 and 45924) are robust to precise fitting choices. We show the version of the fit to 45924 that includes N IV] in the model and mark its wavelength in grey for the other sources.
    At the blue end of PRISM, source 10646 seems resolved into two galaxies (Weaver et al. in prep).
    The blended excess between N III] and Si III] can be well-modeled by Fe II, [Ne III]+Si II, and Al III, which are all present in the \cite{Shen2019} AGN composite \citep{Labbe2024}.
    The remaining UV fits are in Appendix \ref{sec:UV_fits}.}
    \label{fig:example_fits}
\end{figure*}
\begin{figure}
    \centering
    \includegraphics[width=\columnwidth]{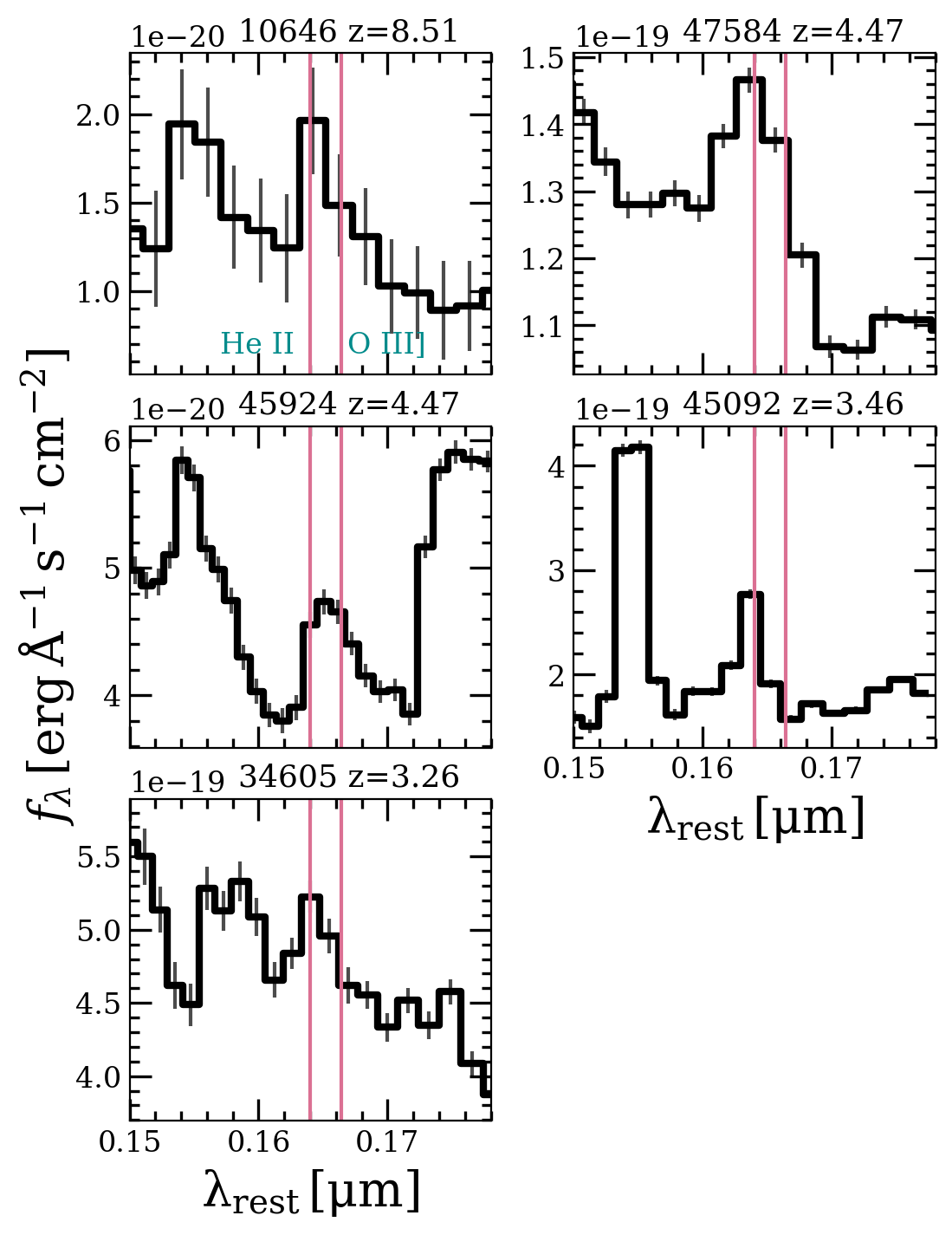}
    \caption{He II + O III] blend for the galaxies with detected He II emission. With the exception of known broad-line AGN 45924, the presence and dominance of He II is visually apparent in each case.}
    \label{fig:zoomed_HeII}
\end{figure}
In this section, we describe our fitting approach for the UV lines used in the \cite{Feltre2016} diagrams (Section \ref{sec:UV_fitting}), H$\beta$ and [O III] (Section \ref{sec:OH_fitting}), 
H$\alpha$ and [N II] (Section \ref{sec:Ha_fitting}),
[O II] and [Ne III] (Section \ref{sec:OH_fitting}), and the weak, high-ionization line [Ne V] (Section \ref{sec:NeV_fitting}). 
We fit the UV lines, H$\beta$ and [O III], and H$\alpha$ and [N II] of the entire parent sample but only fit the other lines for our final sample of galaxies with detected C IV, He II, or C III] emission.

\subsection{Modeling Choices}
\label{sec:choices}
We separately fit five groups of lines, allowing the continuum to be locally well-fit by a power law. We normalize this power law continuum to the blue end of the fitting region and add a Gaussian for each line.
We require the lines within each fitting region to share an intrinsic width. 
Because we only use line ratios for emission lines that are close in wavelength, we do not model dust. Regardless, especially considering that these UV lines are detectable, reddening effects are small compared to our line ratio uncertainties. Likewise, H$\beta$ absorption should be $\lesssim1-2 \mathrm{\mathring{A}}$ \citep[e.g.,][]{GildePaz2003,Gavazzi2004}, which is negligible compared to the EWs and their uncertainties.

We use flat priors on the power law index and amplitude between $-$3 and 3 and 1e$-$22 and 1e$-$18, respectively. For the fits to relatively weak lines (i.e., not for the $\mathrm{H\beta}$+[O III] or $\mathrm{H\alpha}$+[N II] fits), we simultaneously fit for a multiplicative noise inflation term, using an allowed range of 0.1--10. 
A factor of 1.4 is typically a good fit.
To find best fits and corresponding uncertainties, we use the PyMultiNest \citep{Buchner2014} implementation of MultiNest \citep{Feroz2009}. 

Before calculating likelihoods, we convolve the model with the instrument resolution, using the \cite{Jakobsen2022} JDOX pre-launch instrumental broadening.
We adopt a conservative increased resolution of 1.3 \citep[][]{Greene2024}. 
Point sources have better resolution by a factor closer to $\sim$2 \citep{deGraaff2024,Slob2024}, although this level is not preserved through the reduction and co-adding because the pixel size is large relative to the PSF.
We note that lines have a systematic uncertainty on their FWHMs because of our use of this uniform inflation factor rather than a wavelength-dependent, size-dependent factor.
To encapsulate this error, for the final emitters, we repeat the fits without any increase on the resolution. We propagate the resulting line flux differences into our plotted ratio and EW errors. This additional source of uncertainty is negligible (i.e., $<10\%$ additional uncertainty on the line ratios) for all but the highest-SNR sources (45092, 47584, and 45924).
Before convolving the instrumental broadening and the model, we use a wavelength grid that oversamples the variable resolution by a factor of four. We then preserve flux while resampling onto the wavelength grid. In Section \ref{sec:diagnostics}, we present the resulting diagnostic diagrams.

The line fitting presents several challenges: the spectral resolution is both low and variable (R$\sim$50$-$400), and our understanding of the line spread function (LSF) and noise is incomplete, including regarding the magnitude and impact of correlated noise. 
Once correlated noise has been quantified for UNCOVER, fitting should use a covariance matrix (as in JADES et al. in prep).
To explore the robustness of deblending in light of these uncertainties, we test our fitting procedure on a series of mock spectra. We detail these tests in Appendix \ref{sec:simulations} but summarize our main conclusions here.
Namely, we demonstrate that we can satisfactorily deblend lines and represent the corresponding flux uncertainties.
However, we cannot rule out that a couple of our final sources could be false positives. For one source (34605), $\sim1\%$ of mocks with no injected line yield 3$\sigma$ ``detections." We demonstrate that the false positive rates increase modestly when we use correlated noise in the generation of mocks (e.g., with a correlation coefficient of 0.6 between neighboring pixels).

In Figure \ref{fig:example_fits}, we show the spectra and corresponding fits for a few of the best sources in our final sample of rest-frame UV line emitters. 
We also emphasize the unmistakable presence of He II in several sources, regardless of the accuracy of the deblending procedure (Figure \ref{fig:zoomed_HeII}).
In Figures \ref{fig:he2_spectra}--\ref{fig:c45_spectra}, we show the rest-frame UV spectra for the remaining emitters, including marginal detections.

\setlength{\tabcolsep}{2.5pt}

\begin{table*}
\centering
\tiny
\begin{tabular}{ccccccccccccc}
\hline
MSA ID & log(C IV) & log(He II) & log(O III]) & log(N III]) & log(Si III]) & log(C III]) & C IV EW & He II EW & C III] EW & Fig. \ref{fig:ratio_diagnostics} Loc. & log([Ne III]) & log([O II]) \\
\hline
\hline
34605&$<-17.51$&$-17.27^{+0.10}_{-0.17}$&$<-17.08$&$<-17.05$&$<-17.11$&$<-17.10$&$<7$&$2^{+2}_{-2}$&$<9$&AGN&$-16.87^{+0.05}_{-0.06}$&$-16.46^{+0.03}_{-0.03}$\\
45092&$-16.45^{+0.01}_{-0.01}$&$-16.87^{+0.02}_{-0.02}$&$<-18.48$&$-17.52^{+0.07}_{-0.08}$&$-16.94^{+0.03}_{-0.03}$&$-17.13^{+0.14}_{-0.21}$&$49^{+2}_{-2}$&$17^{+1}_{-1}$&$8^{+3}_{-3}$&AGN&$-16.23^{+0.04}_{-0.04}$&$<-19.54$\\
47584&$<-18.29$&$-17.17^{+0.09}_{-0.11}$&$<-17.71$&$-17.72^{+0.07}_{-0.09}$&$<-17.94$&$-17.11^{+0.04}_{-0.05}$&$<2$&$10^{+4}_{-4}$&$15^{+1}_{-1}$&AGN&$-17.01^{+0.03}_{-0.04}$&$-16.91^{+0.03}_{-0.03}$\\
45924&$-17.42^{+0.07}_{-0.09}$&$-18.28^{+0.26}_{-0.74}$&$-18.06^{+0.07}_{-0.08}$&$-17.25^{+0.01}_{-0.01}$&$-17.68^{+0.04}_{-0.04}$&$-17.00^{+0.03}_{-0.03}$&$17^{+8}_{-8}$&$2^{+4}_{-4}$&$45^{+9}_{-9}$&$\mathrm{SFG^{\textbf{*}}}$&$-16.95^{+0.05}_{-0.05}$&$<-19.54$\\
10646 2&$-17.49^{+0.06}_{-0.07}$&$-17.59^{+0.09}_{-0.13}$&$<-17.55$&$<-17.91$&$<-17.56$&$-17.63^{+0.09}_{-0.11}$&$35^{+11}_{-12}$&$27^{+9}_{-10}$&$39^{+12}_{-12}$&AGN&$-17.49^{+0.03}_{-0.03}$&$-17.57^{+0.03}_{-0.03}$\\
42489&$<-17.48$&$<-17.63$&$<-17.71$&$<-17.85$&$<-17.74$&$-17.64^{+0.09}_{-0.13}$&$<18$&$<14$&$18^{+6}_{-6}$&\dots&$-17.70^{+0.04}_{-0.05}$&$-18.01^{+0.09}_{-0.11}$\\
37005&$-17.06^{+0.05}_{-0.05}$&$<-17.61$&$<-17.56$&$<-17.90$&$<-17.75$&$-17.82^{+0.12}_{-0.19}$&$46^{+6}_{-6}$&$<15$&$14^{+5}_{-5}$&\dots&$-17.92^{+0.08}_{-0.10}$&$<-18.10$\\
17467&$<-17.40$&$<-17.33$&$-17.48^{+0.12}_{-0.17}$&$<-17.29$&$<-17.40$&$-17.28^{+0.10}_{-0.15}$&$<8$&$<10$&$14^{+5}_{-5}$&\dots&$-17.11^{+0.04}_{-0.05}$&$-17.29^{+0.06}_{-0.07}$\\
10646 1&$<-17.42$&$<-17.51$&$<-17.45$&$<-17.53$&$<-17.52$&$-17.63^{+0.11}_{-0.18}$&$<17$&$<16$&$18^{+4}_{-4}$&\dots&$-17.49^{+0.03}_{-0.03}$&$-17.57^{+0.03}_{-0.03}$\\
10155&$-17.87^{+0.06}_{-0.09}$&$<-17.83$&$<-18.15$&$<-17.88$&$<-17.96$&$<-18.00$&$52^{+10}_{-12}$&$<71$&$<67$&\dots&$-18.57^{+0.10}_{-0.13}$&$<-18.44$\\
36755&$-17.73^{+0.06}_{-0.06}$&$<-17.93$&$<-17.89$&$<-18.06$&$<-18.09$&$<-18.12$&$88^{+18}_{-15}$&$<66$&$<56$&\dots&$<-18.63$&$<-18.87$\\
23604&$-17.95^{+0.09}_{-0.12}$&$<-18.14$&$<-18.17$&$<-17.97$&$<-18.16$&$<-18.09$&$51^{+14}_{-14}$&$<38$&$<60$&\dots&$<-18.54$&$<-19.57$\\
\hline 
\end{tabular}
\caption{Flux measurements ($\mathrm{erg \, s^{-1} \, cm^{-2}}$) and rest-frame EWs ($\mathrm{\mathring{A}}$) for lines used in the UV diagnostic diagrams for the emitter sample. 
As we discuss in Section \ref{sec:choices}, the final uncertainties include a term for the difference between the MultiNest results using the pre-launch instrumental broadening and an increased resolution of 1.3. This extra term is significant (but conservative) for the three highest-SNR sources (45092, 47584, and 45924). 
We note the classifications of the galaxies according to their consistency with the AGN and SFG models in Fig. \ref{fig:ratio_diagnostics}, but all emitters without He II detections are consistent with both model grids. We place an asterisk next to the SFG designation for BLAGN 45924, because it is typical to see broad line ratios falling in the SFG region of the diagnostics (Section \ref{sec:sf_literature}, Figure \ref{fig:literature_diagnostic}).}
\label{tab:emitters}
\end{table*}

\subsection{UV Lines}
\label{sec:UV_fitting}
First, we fit C IV $\lambda \lambda$1548, 1551, O III] $\lambda \lambda$1661, 1666, N III] $\lambda$1750, [Si III] $\lambda$1883+Si III] $\lambda$1892, and C III] $\lambda \lambda$1907, 1909, with the fitting region extending to 0.02 rest-frame microns blueward of C IV and redward of C III]. 
Historically, N V $\lambda$1240 has also been a key AGN indicator \citep{Laporte2017,Jaskot2017,Vanzella2018,Mignoli2019,Matthee2022}, but it is impossible to isolate with PRISM.
For prior ranges on line fluxes, we use from 1e$-$21 $\mathrm{erg \, s^{-1} \, cm^{-2}}$ to double the \texttt{LMFIT} \citep{Newville2016} maximum flux when fitting a power-law continuum plus C IV, He II, and C III]. 
We require all the lines to share a FWHM between 50 and 1000 km/s. 
Although the spectral resolution prevents us from constraining the line width, this range allows the fit to marginalize over this source of uncertainty.
Because of the blending, we measure the EWs from the model fits.

Although we fix the redshifts for these initial fits to the overall solutions, we iterate on the fits for the final sample by freeing $z$ by 0.01 in each direction. This change allows a marginalization over wavelength calibration uncertainty.
Three emitters (45924, 47584, 45092) have sufficiently high SNR to remove the noise inflation parameter in this updated fit.
We therefore use the parameters from these free-redshift fits for the diagnostic plots and include the fluxes and EWs in Table \ref{tab:emitters}. 

\subsection{H$\beta$ and [O III]}
\label{sec:OH_fitting}
We also fit H$\beta$ and the [O III] doublet for the entire parent sample, with the doublet flux ratio fixed to 3.1 \citep{Osterbrock1989}. The fitting region includes 0.06 microns beyond the lines and we fix the redshift to the overall solution for the source. 
\texttt{LMFIT} once again provides an allowed range of values in the MultiNest fit.
We only fit a shared narrow component to the lines. For the LRDs, which have broad H$\beta$, we use the \cite{Greene2024} fits to account for both the narrow and broad components.

\subsection{H$\alpha$ and [N II]}
\label{sec:Ha_fitting}
To check for evidence of broad H$\alpha$, we fit H$\alpha$ and the [N II] doublet, with a doublet ratio of 3.05 and the redshift allowed to vary by 0.01 in each direction. 
Following \cite{Greene2024},
for the two-component fit to be preferred over the narrow-only one, we require a 3$\sigma$ improvement on $\chi^{2}$ and a 5$\sigma$ detection of the broad component by MultiNest. 
At PRISM resolution, we are only sensitive to $\gtrsim2000 \, \rm km/s$ broad lines.
While there are no new broad-line AGN in our final emitter sample, we report one non-LRD BLAGN, which has evidence for UV lines at the $\sim$2.5$\sigma$ level (Figure \ref{fig:11254}).

\subsection{[Ne V]}
\label{sec:NeV_fitting}
We fit the high-ionization line [Ne V] $\lambda$3426 for the objects in the final line emitter sample.
Because [Ne V] is situated in a complicated part of the continuum, we fit 0.04 microns (rest-frame) blueward of [Ne V] through 0.025 microns redward of the line, include He II $\lambda$3204 in the model, and fix the redshift to the overall solution.
Although the continuum near [Ne V] is not well-described by a power law, these choices allow for reasonable fits (Figure \ref{fig:example_fits}) and the two detections (45092 and 45924) are robust to precise fitting choices.

\subsection{[O II] and [Ne III]}
\label{sec:ON_fitting}
In our fits to [O II] $\lambda$3726 and [Ne III] $\lambda$3869, we use the continuum 0.04 microns blueward of [O II] and redward of [Ne III]. The model also includes H$\epsilon$ and H$\delta$ since they sit so close to [Ne III].

\begin{figure}
    \centering
    \includegraphics[width=\columnwidth]{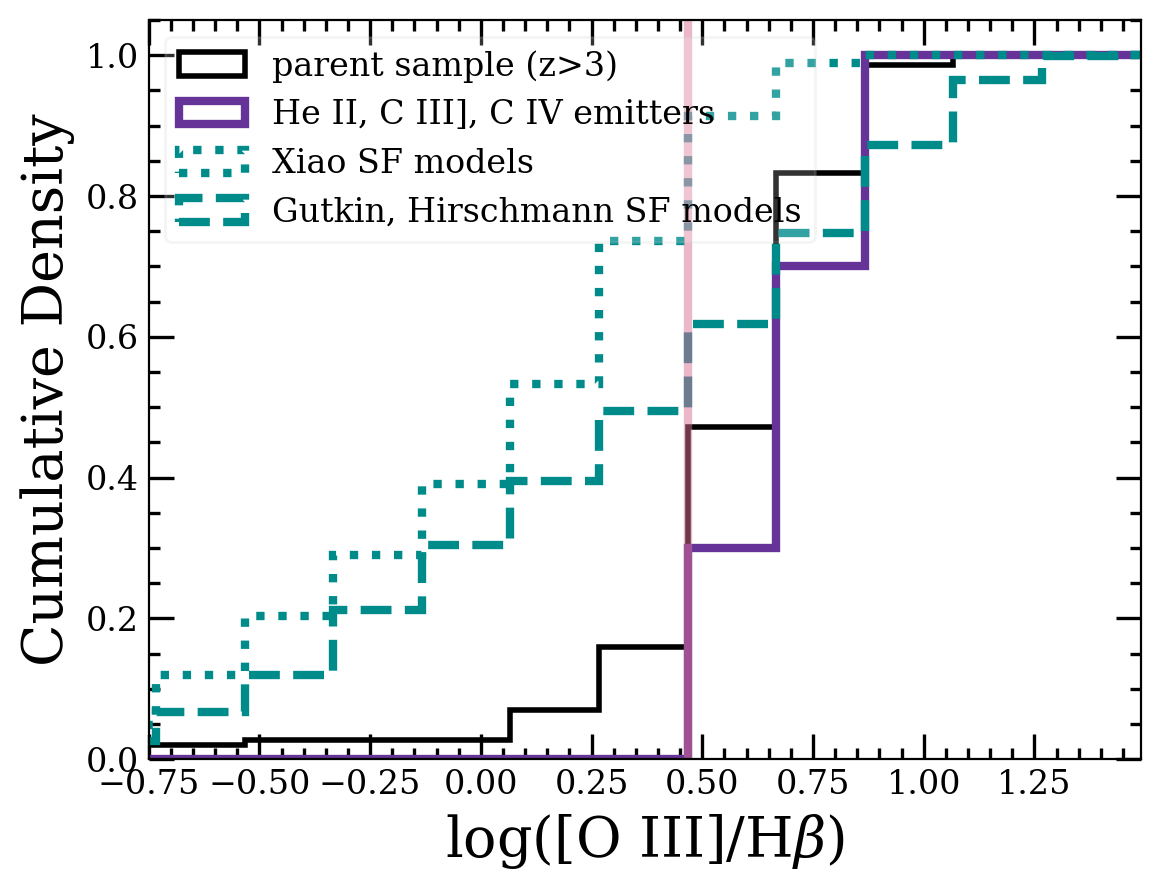}
    \caption{Cumulative distribution functions of log([O III]/H$\beta$) for the parent sample, the emitters, and the model grids for star-forming galaxies \citep{Gutkin2016,Xiao2018,Hirschmann2019}. The vertical pink line marks the minimum value measured among the final line emitters, which we use to cut down on the model grids (Section \ref{sec:o3hb}).}
    \label{fig:O3Hb_hist}
\end{figure}
\begin{figure*}
    \centering
    \includegraphics[width=2\columnwidth]{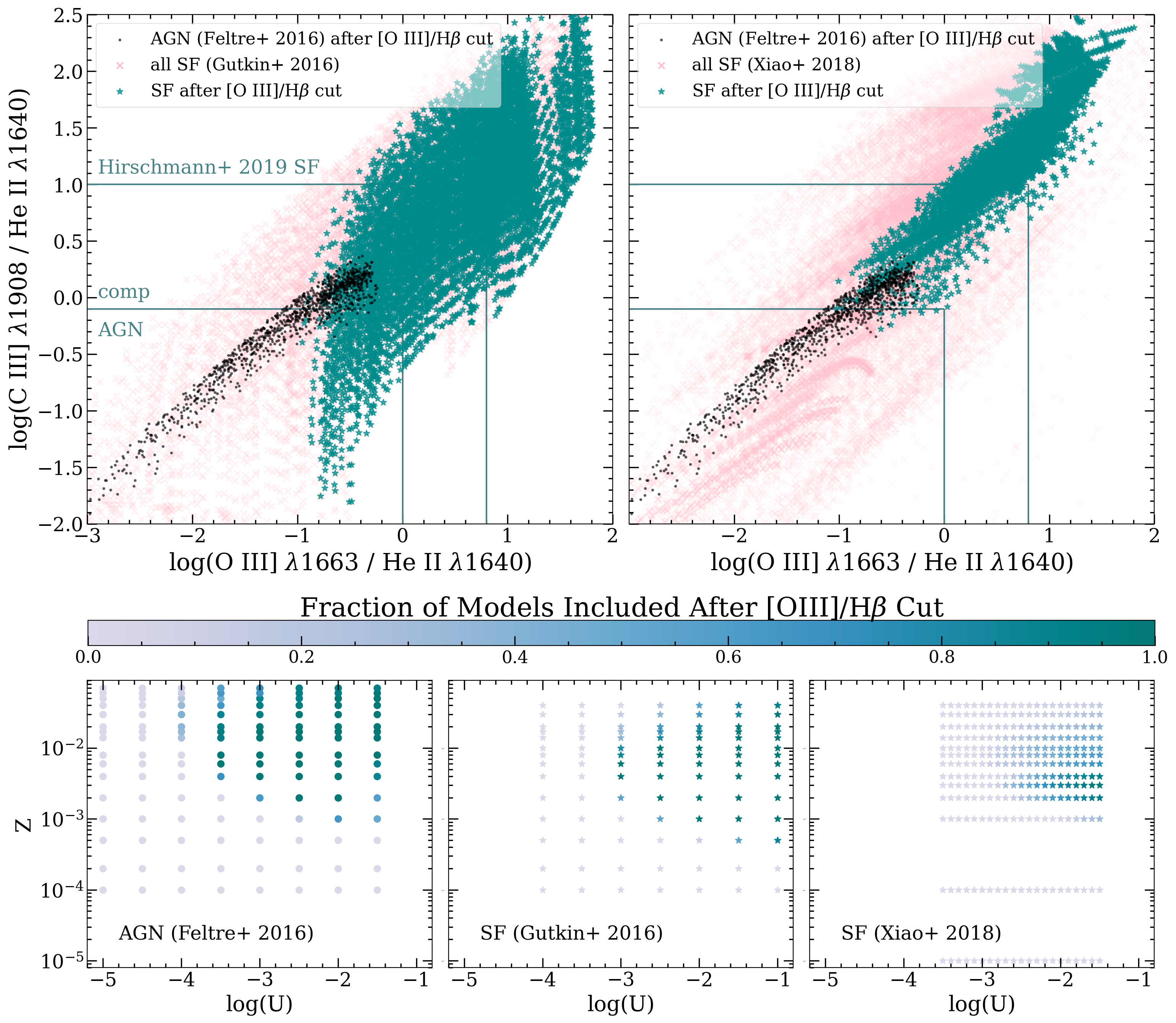}
    \caption{An example of the effect of the [O III]/H$\beta$ cut on the models relevant to our line-emitting sample in the \cite{Feltre2016} diagnostics. In the top row, we plot post-cut \cite{Feltre2016} AGN models in each panel and note that this cut does not have a significant impact in this part of parameter space. On the other hand, the \cite{Gutkin2016} (left) and \cite{Xiao2018} (right) star-forming galaxy models are clearly constrained when we remove models with [O III]/H$\beta$ values inconsistent with those we find in the sample of He II or C III]-emitters (Section \ref{sec:o3hb}). In particular, while the full model grids (pink X's) have extensive overlap with the AGN models, the allowed models (cyan stars) are largely restricted to the \cite{Hirschmann2019} star-forming and composite regions. 
    In the bottom row, we show the metallicity and ionization parameter constraints for each model grid, by coloring the grid points in that space by the fraction of models still included after the [O III]/H$\beta$ cut.} 
    \label{fig:O3Hb_cut}
\end{figure*}
\begin{figure*}
    \centering
    \includegraphics[width=2\columnwidth]{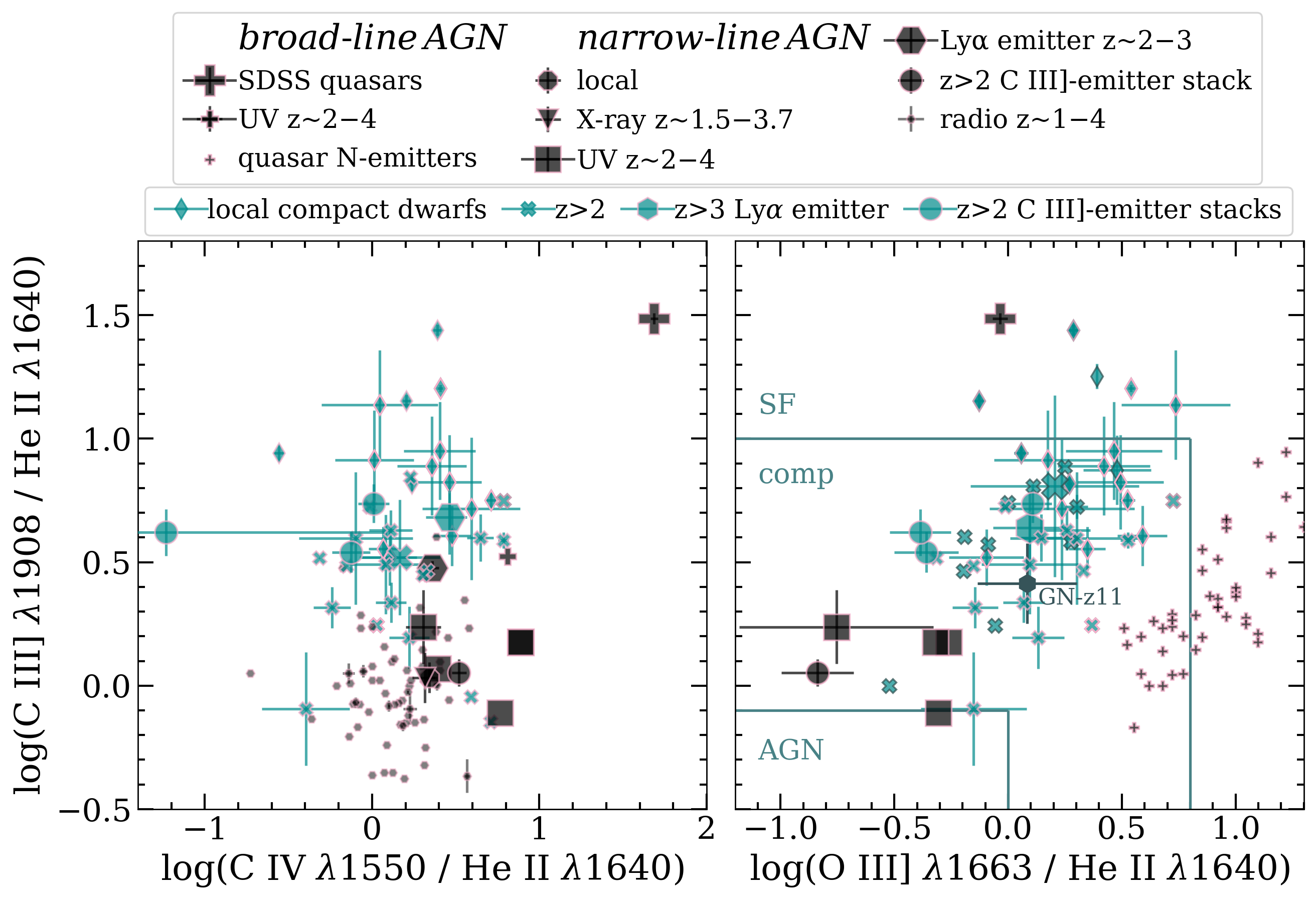}
    \caption{Literature source locations on two of the UV ratio-ratio diagnostics.
    The panels include AGN and star-forming galaxies, with AGN contribution ruled out to varying degrees. We outline markers in pink if the galaxy has C IV detected in emission (Section \ref{sec:candidacy}). Larger points are composites and smaller ones mark individual galaxies.
    For AGN, we include the SDSS quasar composite \citep{VandenBerk2001}, C IV-selected BLAGN \citep{Mignoli2019}, individual SDSS quasars with N IV] and N III] emission \citep{Batra2014}, UV-selected NLAGN at cosmic noon \citep{Hainline2011,Alexandroff2013,Mignoli2019}, local optically-selected NLAGN \citep{Nagao2006}, X-ray AGN \citep{Nagao2006}, Ly$a$-emitting AGN at cosmic noon \citep{Sobral2018}, AGN from a sample of C III] emitters \citep{LeFevre2019}, and radio galaxies \citep{DeBreuck2000}.
    For SFGs, we include individual local compact dwarf galaxies \citep{Berg2016,Berg2019,Senchyna2017,Senchyna2019}, higher-redshift individuals \citep{Amorin2017,Erb2010,Fosbury2003,Patricio2016,Saxena2020,Schmidt2021,Topping2024,Vanzella2020} and composites \citep{Steidel2016,Tang2023}, and stacks of Ly$a$ emitters \citep{Sobral2018,Feltre2020} and C III] emitters \citep{LeFevre2019}.
    We also include the ambiguous source GN-z11 \citep[green hexagon,][]{Bunker2023}.
    \cite{Hirschmann2019} optimized the plotted cutoffs with the purity and completeness fractions of different $z\sim0-6$ ionization sources in post-processed cosmological simulations.} 
    \label{fig:literature_diagnostic}
\end{figure*}
\section{UV and Optical Diagnostic Diagrams}
\label{sec:diagram_intro}
Now that we have the line strengths in hand, we provide an overview of the models that inform our use of UV emission-line ratios to classify galaxies.
A common approach to identifying AGN is the use of ratios between nearby emission lines. These ratios probe the shape of the ionizing spectrum, which we expect to be harder in AGN than in star-forming galaxies (SFGs). For instance, in the local Universe, samples of AGN have been successfully selected at high values of [O III]/H$\beta$ and [N II]/H$\alpha$ in so-called BPT diagrams \citep{Baldwin1981,Veilleux1987,Kauffmann2003,Kewley2006}. 

However, properties of the ISM also affect these line ratios. Modeling and observations both indicate that optical line ratios fail to separate AGN and SFGs beyond z$\sim$1.5 because SFGs can move up to the AGN region and metal-poor AGN move towards the SF region of the diagrams \cite[e.g.,][]{Shapley2005,Groves2006,Kewley2013o,Kewley2013t,Juneau2014,Steidel2014,Nakajima2022}. This redshift evolution is driven by some combination of metallicity, abundance patterns, ionization parameter, sSFR, and electron density, all of which are significantly different at cosmic noon and higher redshifts \citep[e.g.,][]{Brinchmann2008BPT,Sanders2015,Sanders2016,Hirschmann2017,Strom2017,Strom2018,Curti2024,Tacchella2024}.

Nevertheless, the same models that predict the failure of optical line ratio diagrams reveal promising line ratios in the UV \citep[e.g.,][]{Feltre2016,Hirschmann2023}. 
It is also convenient to use UV ratios for high-redshift galaxies as rest-frame optical lines get pushed deeper into the IR. Moreover, some individual UV lines might be sufficient for AGN selection if the ionization potential requires an AGN.

In this section, we introduce the diagnostics we use to separate high-redshift AGN candidates from interesting star-forming galaxies. In Section \ref{sec:models}, we provide an overview of these models as well as the simple cutoffs from \cite{Hirschmann2019}. 
In Section \ref{sec:salvaging}, we leverage strong optical lines to limit the allowed model space and note the locations of literature AGN and SFGs with the necessary line detections. At the same time, these data underscore challenges on the modeling side, such as the inability to reproduce observed UV EWs in galaxies that do not seem to host AGN. 
Putting this together, we conclude this section with a prescription for AGN candidacy (Section \ref{sec:candidacy}). 
In Section \ref{sec:diagnostics}, we then place our objects on ratio and EW diagnostics and note other relevant features of their spectra.

\subsection{Models}
\label{sec:models}
There are a few UV diagnostics that we explore in this work, but, for the following reasons, we focus on the \cite{Feltre2016} UV diagnostic diagrams, which involve the ratios of He II $\lambda$1640 with each of C IV $\lambda \lambda$1548, 1551, O III] $\lambda \lambda$1661, 1666, N III] $\lambda$1750, [Si III] $\lambda$1883+Si III] $\lambda$1892, and C III] $\lambda \lambda$1907,1909. 

A proposed selection method that incorporates both UV and optical lines is the [O III]/H$\beta$ vs. [Ne III]/[O II] (OHNO) diagram \citep{Zeimann2015,Backhaus2022,Larson2023,Cleri2023}. However, the AGN region is contaminated with SFGs even at low-redshift \citep{Backhaus2022} and sub-solar metallicities complicate both of the ratios in the OHNO diagram \citep[e.g.,][]{Trouille2011,Scholtz2023}. Nevertheless, some AGN may have ratios so extreme that they are indeed isolated in this space \citep[e.g.,][]{Zeimann2015,Larson2023}.

Individual lines can also serve as AGN indicators when their ionization potential is so high that stars are unlikely to account for any detection. The mere presence of [Ne V] in a galaxy spectrum is often considered sufficient evidence for an AGN, thanks to the ionization potential of 97 eV \citep[e.g.,][]{Schmitt1998,Gilli2010,Mignoli2013}, but there are cases of detected [Ne V] from seemingly inactive galaxies \citep[e.g.,][]{Izotov2004,Izotov2012,Izotov2021}. Especially at high redshift or low metallicity, [Ne V] may suffer from the same degeneracies as other UV diagnostics. However, the pairing of high-ionization lines with other features (e.g., [Ne III]) helps to distinguish between AGN and other ionization sources \citep[e.g., shocks or Pop III stars,][]{Abel2008,Izotov2021,Cleri2023,Chisholm2024,Silcock2024}. Unfortunately, even in AGN, [Ne V] is intrinsically weak and highly reddened \citep[e.g.,][]{Netzer1990}, so we cannot rely on it for AGN selection, especially at low spectral resolution.

On the other hand, the \cite{Feltre2016} diagnostics are promising, even at the PRISM resolution. To identify these useful line ratios, \cite{Feltre2016} presented AGN models that use CLOUDY \citep{Ferland1993,Ferland1998,Ferland2013} and are parameterized by hydrogen density, ionization parameter (i.e., the ratio between the number densities of Hydrogen-ionizing photons and Hydrogen), metallicity, 
mass ratio of dust and metals, and the power law index of the ionizing spectrum. After pairing the AGN grids with the \cite{Gutkin2016} stellar models, \cite{Feltre2016} found that the key diagnostics leverage the expectation that relatively high values of He II $\lambda$1640 signal contribution from an AGN narrow-line region \citep[e.g.,][]{Villar-Martin1997,Allen1998}. The success of He II as a tracer is linked to its recombination origin (with an ionization energy of 54.4 eV), whereas the nearby lines are largely collisionally-excited metal lines.
We use all of the effective line ratios presented in \cite{Feltre2016} but emphasize the diagrams with C III]/He II vs. C IV/He II and C III]/He II vs. O III]/He II (e.g., Figure \ref{fig:literature_diagnostic}) because they include the lines that are typically easier to detect. 

More recent studies have improved these models. As part of an effort to couple nebular emission lines to cosmological zoom-in hydrodynamical simulations, \cite{Hirschmann2019} included composite galaxies with both stellar and AGN contributions. By optimizing purity and completeness, they identified cutoffs between the AGN, composite, and SFG regions of the diagnostics, where the composite region includes galaxy models with a ratio of $10^{-4}-10^{-2}$ between black hole accretion rate and star formation rate. Since they incorporated updated spectra of Wolf-Rayet stars, we use their versions of the \cite{Gutkin2016} models in this work.

Still absent from these models, though, are rotation and binary products, which seem essential for understanding UV emission from star-forming galaxies \citep[e.g.,][]{Eldridge2012}.
To address this issue, \cite{Xiao2018} used CLOUDY but with BPASS\footnote{https://bpass.auckland.ac.nz/index.html} models \citep{Stanway2018}. Crucially, they found that the resulting stellar model grids span the entire diagnostic space, potentially rendering the UV line ratios unhelpful after all. However, we demonstrate that the He II diagnostics remain empirically relevant for two reasons, as we discuss in Section \ref{sec:salvaging}.

Still, there are clear caveats to the models used herein. While the grids generally include wide parameter ranges, they mainly stick to solar abundance ratios
and the \cite{Feltre2016} AGN models only scale the ionizing spectrum with accretion luminosity.
The currently inevitable uncertainty, though, is that there are few observations with which to check and calibrate the modeled UV emission \citep[e.g.,][]{Whitler2023}. For instance, metal-poor models (i.e., $\lesssim20\% \, Z_{\odot}$) have not been sufficiently tested with observations since there are limited studies of resolved lower-Z O stars \citep[e.g.,][]{Telford2023}. 
\subsection{Semi-Empirical Diagnostic Boundaries}
\label{sec:salvaging}
\subsubsection{[O III]/H$\beta$ Limits Model Grid Overlap}
\label{sec:o3hb}
There is significant overlap between the full AGN and SFG model grids, but we can directly cut down on the grid parameters using other information offered by the PRISM wavelength coverage.
In particular, after cutting out models with [O III]/H$\beta$ values inconsistent with the distribution within our final sample, we see renewed support for the usefulness of the diagnostics.

At these redshifts, we no longer trust AGN selection methods hinging on optical ratios like [O III]/H$\beta$, because of the sensitivity to not just the ionizing spectrum but to the ionization parameter and metallicity. However, the sensitivity to these parameters allows [O III]/H$\beta$ to constrain them and thus to decrease the degeneracy between possible origins of the observed UV line ratios.

We plot the [O III]/H$\beta$ distributions for our parent sample, our line emitters, and the SFG model grids in Figure \ref{fig:O3Hb_hist}. 
Using the minimum [O III]/H$\beta$ of 3.0 found among our emitters, we apply a cut to the model grids.
In Figure \ref{fig:O3Hb_cut} we show the model diagnostic diagram before and after applying this cut. The cut drastically reduces the span of the restricted SFG model grid, such that there is limited overlap between the AGN and SFG models.

For each metallicity and ionization parameter combination, we also plot the fraction of models still included after the cut.
Both extremes of metallicity in the model grids are then disfavored or entirely removed, with the highest fraction of models kept at Z$\sim$0.005. The effect on ionization parameter, on the other hand, is monotonic: for the stellar models, values of logU lower than $-3.2$ cannot survive this cut, and the cut's effect decreases when moving to higher ionization parameters in the grid. Interestingly, although the full \cite{Xiao2018} model grid extends more than the \cite{Gutkin2016} one into the composite and AGN regions, it occupies less parameter space than the \cite{Gutkin2016} models once the cut has been applied to both. We adopt the updated \cite{Gutkin2016} models in the plotted diagnostics to show the more conservative span of SFG models.

\subsubsection{Observed Line Ratio Diagram Samples from the Literature}
\label{sec:sf_literature}
Literature objects also support the viability of the UV diagnostics.
In Figure \ref{fig:literature_diagnostic}, we place literature AGN and SFGs on the two most important ratio diagnostics (C III]/He II vs. C IV/He II and C III]/He II vs. O III]/He II). Here we explain the basics of the included objects and highlight key take-aways that affect how we interpret the locations of the UNCOVER galaxies. In short, most literature AGN and SFGs with the necessary line detections live in the \cite{Hirschmann2019} composite region, but narrow-line AGN and SFGs occupy different areas within that region. 

First, we note that these diagnostics only apply to the narrow-line region of AGN. The \cite{Feltre2016} AGN models do not consider the temperature and density of the broad-line region and we see that the broad lines of SDSS quasars are deep in the star-forming region. We show this discrepancy using both the SDSS composite \citep{VandenBerk2001} and the subset of nitrogen emitters \citep{Batra2014}, as well as with the composite of the subset of \cite{Mignoli2019} C IV-selected AGN that exhibit broad lines.
At PRISM resolution in the rest-frame UV, we generally cannot distinguish broad from narrow lines, so we do not expect to select any AGN with UV features dominated by the broad-line region.

Second, the diagnostic works for a variety of narrow-line AGN populations (UV-selected at cosmic noon: \citealt{Hainline2011,Alexandroff2013,Mignoli2019}, optically-selected locally: \citealt{Nagao2006}, X-ray-selected: \citealt{Nagao2006}, selected within a sample of C III] emitters: \citealt{LeFevre2019}, and radio galaxies: \citealt{DeBreuck2000}), but AGN often live in an ambiguous region with model overlap or even just at the edge of the SFG models.
Similarly, the star-forming galaxies tend to be consistent with models, but with most being in the \cite{Hirschmann2019} composite region. We include $z>2$ star-forming galaxies and stacks as well as their local low-metallicity dwarf analogues. Despite the benefit of the low-redshift analogues for probing poorly-understood low-metallicity star formation, the few seemingly star-forming objects sitting among the AGN in Figure \ref{fig:literature_diagnostic} are all higher-redshift.
However, for the star-forming galaxy most consistent with the literature AGN in the log(C III]/He II) vs. log(O III]/He II) diagnostic, the C IV detection is marginal and it is difficult to rule out contribution from an AGN \citep{Amorin2017}.
Because of these ambiguities and the PRISM spectral resolution, we focus on identifying the best AGN candidates, noting that these objects would benefit from higher-resolution follow-up.
\subsection{AGN Candidacy}
\label{sec:candidacy}
We have seen that the (post-cut) models and literature sources support the utility of the UV diagnostic diagrams for the identification of narrow-line AGN candidates. Still, there is significant overlap in the diagrams thanks to different possible physical causes of the ionization in the rest-frame UV.

To help break the remaining degeneracy, the existing data suggest we can leverage C IV information and EWs. Stellar systems with relatively strong He II, which could be misinterpreted as AGN in the UV diagnostics, often have a broad He II component and a P Cygni or weak C IV profile \citep[][]{Conti1996,Leitherer1996,Shapley2003,Erb2010,Sobral2015,Nanayakkara2019,Saxena2020}.
This combination of features is usually attributed to Wolf-Rayet stars, unless the EWs are above $\sim$3$\mathrm{\mathring{A}}$, in which case Very Massive Stars ($M>100\mathrm{M_{\odot}}$) are invoked \citep[e.g.,][]{Grafener2015,Martins2023,Wofford2023,Upadhyaya2024}.
Even with this absence of velocity information, we can be more confident in AGN candidates if they have C IV emission.

EW measurements present another potential way to discriminate between different ionization mechanisms. In the C III] EW vs. C III/He II and C IV EW vs. C IV/He II planes, \cite{Nakajima2018} suggested cutoffs in which high EW alone can require an AGN. For example, their modeling indicates that objects with C III] EW above 20$\mathrm{\mathring{A}}$ must be AGN.
Complicating this approach is the fact that star formation models fail to account for the highest observed EWs of C IV, He II, and the nearby metal lines \citep[e.g.,][]{Shirazi2012,Jaskot2016,Steidel2016,Senchyna2017,Berg2018,Nanayakkara2019,Plat2019,Olivier2022,Saxena2020,Saxena2022,Stanway2019}. Particularly because this problem is clearest at low metallicity \citep[e.g.,][]{Plat2019}, there is no consensus as to whether AGN alone bridge this gap \citep[e.g.,][]{Shirazi2012}.
The only AGN-free models that have replicated observations require bursty star formation histories, X-ray binaries, radiative shocks, and a top-heavy IMF \citep[][]{Lecroq2024}.

Thus, degeneracy in these diagnostics will persist at least until we have better emission-line models for a variety of ISM conditions.
At our spectral resolution, we can therefore require our most confident AGN candidates to be consistent with the AGN models in the line ratio diagnostics and to have C IV detected in emission. We can further boost confidence in an AGN classification for cases with relatively high EWs, but it remains to be seen the extent to which such EWs can be modeled by other processes. Still, especially as photoionization modeling evolves to describe larger data sets, we cannot rule out that other objects may also host AGN. 
\begin{figure*}
    \centering
    \includegraphics[width=2\columnwidth]{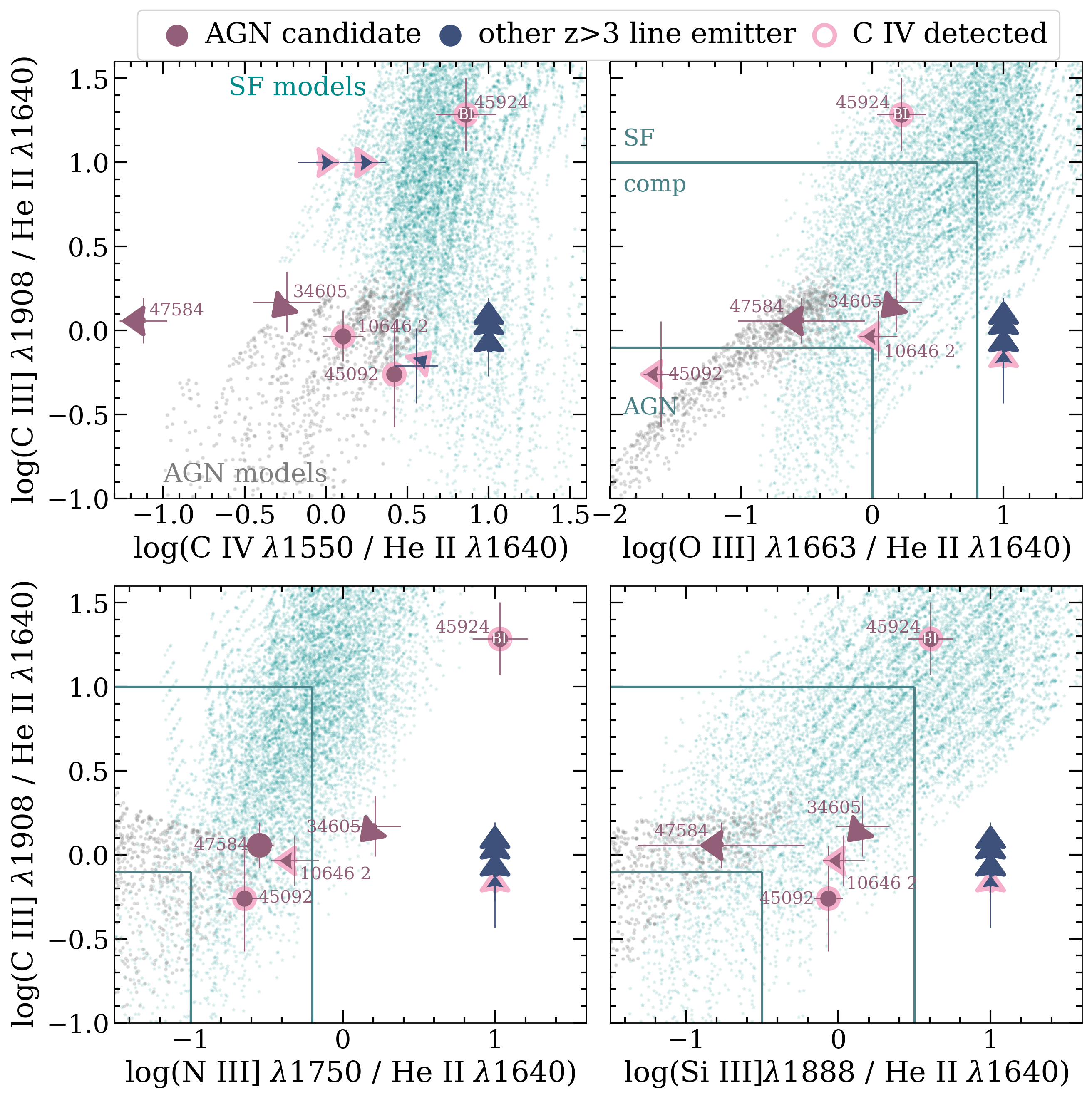}
    \caption{UV ratio-ratio diagnostics for our sample of UNCOVER sources with C III] or He II detected above 3$\sigma$ or C IV detected above 5$\sigma$. We outline points in pink if they have C IV detected (Section \ref{sec:candidacy}). The ``AGN candidates" are the sources with He II detections, which are BLAGN 45924 and the ones most consistent with the AGN models in the top left panel. We label these points and limits with their MSA IDs in each panel. As expected based on the location of the SDSS quasars in Figure \ref{fig:literature_diagnostic}, the BLAGN 45924 does not sit on the AGN model grids (Section \ref{sec:sf_literature}, Figure \ref{fig:literature_diagnostic}). 
    We use the same color for the AGN candidates in the remaining figures.
    Arrows indicate 3$\sigma$ upper limits for undetected lines.
    Sources with only C III] above 3$\sigma$ are marked at an x-axis ratio of 1 in each diagram. Similarly, sources with only C IV detected are marked at a C III]/He II ratio of 1 in the top left panel. 
    The cutoffs between AGN, composites, and star-forming galaxies are from \cite{Hirschmann2019} and in black points and cyan stars we mark the \cite{Feltre2016} and \cite{Gutkin2016} AGN and SFG model grids with the [O III]/H$\beta$ cut imposed (Section \ref{sec:o3hb}).}
    \label{fig:ratio_diagnostics}
\end{figure*}
\section{Diagnostics with UNCOVER UV Line Emitters}
\label{sec:diagnostics}
In this section, we introduce the sample of UV line emitters and identify the subset that we consider AGN candidates based on their location in the UV diagnostic diagrams. To do so, we place these objects on He II-based ratio (Section \ref{sec:ratio-ratio}) and EW-ratio diagnostics (Section \ref{sec:EW-ratio}). 
We discuss evidence for broad H$\alpha$ in one non-LRD source (Section \ref{sec:broad_Ha}) and then highlight the potential presence of [Ne V] in a couple of sources (Section \ref{sec:neon_detections}). We also place this sample on the OHNO diagram (Section \ref{sec:OHNO}). 
These efforts divide the sample of line emitters into AGN candidates and ambiguous sources that may be AGN or exceptional star-formers (Section \ref{sec:characterization}).

\subsection{UV Ratio-Ratio Diagnostics}
\label{sec:ratio-ratio}
Our final sample of UV line emitters consists of the galaxies with a detection above 5$\sigma$ for C IV or above 3$\sigma$ for He II or C III].
We choose these lines since He II is in every ratio in the diagnostics, C III] is in every y-axis ratio, and C IV (non-)detections are of particular note as we try to separate sources of ionization (Section \ref{sec:sf_literature}). 
Since C IV is unblended, it is easier for a noise feature to be deemed significant at the 3$\sigma$ level; a comparable feature at the location of He II or C III] would split flux with O III] or Si III] in the fit. We therefore use the higher threshold for C IV detections, finding that our suite of mocks also supports this choice (Section \ref{sec:simulations}).
In Figures \ref{fig:he2_spectra}--\ref{fig:c45_spectra}, we show the fixed-redshift and free-redshift fits to this final sample of line emitters. We use the free-redshift fits for the diagnostics to marginalize over uncertainty in the redshift and wavelength solution.

From the parent sample of \parentnum~$z>3$ galaxies, we have five He II emitters, eight C III] emitters (four of which are also He II emitters), 
and seven C IV emitters. Of the C IV emitters, three have only C IV detected. 

In Figure \ref{fig:ratio_diagnostics}, we place these line emitters on four diagnostics. Each diagnostic involves three lines, requiring most sources to be plotted with 3$\sigma$ limits. In cases with only C IV or C III] detected, we fix the limits at a y- or x-axis value of one, respectively. These C IV and C III] emitters without He II detected are consistent with both the AGN and SFG model grids. We therefore focus our discussion of AGN candidates on the He II emitters, which, apart from 45924, are more consistent with the AGN models in the C III]/He II vs. C IV/He II diagram and in an ambiguous region in the remaining diagrams. 
Because of the locations of literature AGN in these diagnostics (Figure \ref{sec:sf_literature}), we do not require our candidates to sit in the \cite{Hirschmann2019} AGN region.

Using the MSA IDs, we label the AGN candidates. LRD 45924 is deep in the SFG model region, as is expected for broad-line AGN (Section \ref{sec:sf_literature}). One candidate (47584) has particularly low C IV/He II, making it inconsistent with both model grids. A complicated underlying C IV profile (i.e., with P-Cygni, nebular, and resonant contributions) may drive this discrepancy with the models, since the PRISM spectral resolution requires us to fit all of these components with a single Gaussian.

The list of narrow-line AGN candidates includes one of the two seemingly resolved (Weaver et al. in prep) galaxies in an ionization bubble at $z=8.5$ \citep[10646,][]{Fujimoto2023bubble}, a source at a very similar redshift to LRD 45924 (47584), the most massive galaxy in the parent sample (45092, Bezanson et al. in prep), and the one He II emitter without C III] also detected (34605). 
In Appendix \ref{sec:UV_fits}, we comment on interesting aspects of each of these sources and of the broad-line candidates.

\subsection{UV EW-Ratio Diagnostics}
\label{sec:EW-ratio}
Similar EW-based diagnostics are helpful since they only examine two emission lines at a time and allow us to compare the EWs to those of literature AGN and star-formers.

\begin{figure*}
    \centering
    \includegraphics[width=2\columnwidth]{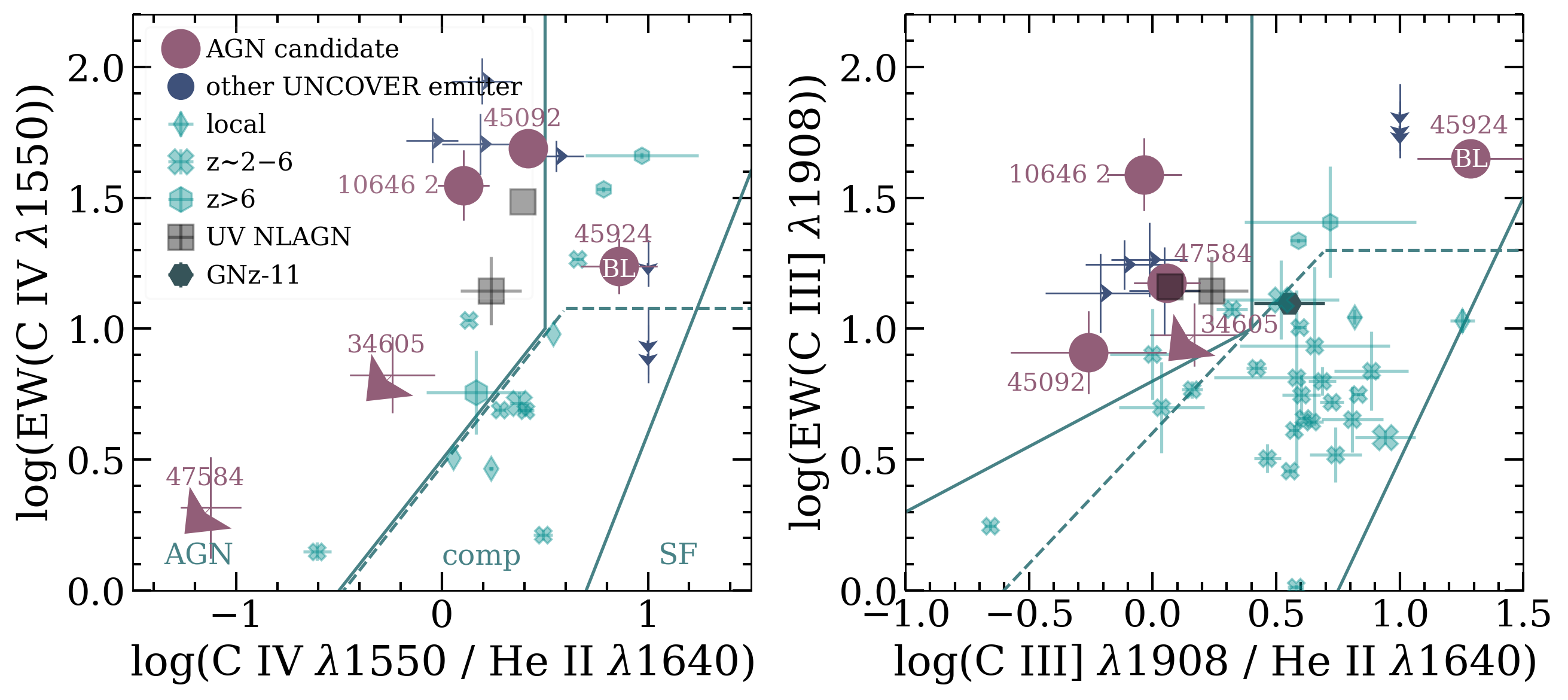}
    \caption{EW-based diagnostics with literature sources and our line emitters marked. AGN candidates from the C III]/He II vs. C IV/He II diagnostic (Figure \ref{fig:ratio_diagnostics}) and from broad H$\alpha$ are larger and in red.
    Solid lines mark the \cite{Hirschmann2019} cutoffs with AGN on the left, composites in the middle, and SFGs on the right.
    Dashed lines follow the \cite{Nakajima2018} cutoff between AGN (left and high) and SFGs (right and low).
    The black square shows the location of stacked UV-selected narrow-line AGN \citep{Hainline2011,Mignoli2019} and the cyan points mark SFGs in three redshift bins: ``local" \citep{Olivier2022,Senchyna2019}, $z\sim2-6$ \citep{Erb2010,Du2020,Marques-Chaves2020,Patricio2016,Schmidt2021,Saxena2020,Saxena2022}, and $z>6$ \citep{Castellano2024,Tang2023,Topping2024}. Also marked in the right panel is GN-z11 \citep[dark green hexagon,][]{Bunker2023}; we exclude it from the left panel because of its P-Cygni C IV profile.
    Sources with neither line detected have EW limits shown at an x-axis ratio of 1.}
    \label{fig:EW_diagnostics}
\end{figure*}

We include the C IV- and C III]-based EW diagnostics in Figure \ref{fig:EW_diagnostics}. We mark the \cite{Hirschmann2019} cutoffs between AGN, composites, and SFGs, as well as the \cite{Nakajima2018} version, which is just between AGN and SFGs.
The \cite{Nakajima2018} separation requires anything at relatively high EW to have an AGN contribution, and thus places in the AGN region supposed high-redshift, extreme star-formers observed with JWST \citep{Topping2024,Castellano2024}. 
Extreme UV EWs are not restricted to a couple outlier high-redshift galaxies; even composite PRISM spectra show a correlation between C III] EW and redshift \citep{Roberts-Borsani2024}.
Thus, the included literature populations are more consistent with the \cite{Hirschmann2019} cutoff than with the \cite{Nakajima2018} one.

Turning to the locations of the objects in our sample within the diagnostic plots, we see that besides 45924, the He II emitters are consistent with the AGN region, providing important additional support to the findings from Figure \ref{fig:ratio_diagnostics}. 
Still, ambiguity remains, at least for the objects with both lines detected, thanks to the proximity to some of the literature galaxies deemed SFGs.

Furthermore, we present a few C IV and CIII] emitters (45092, 45924, 36755, 37005, and 23604) with comparable EWs to those observed in exceptional $z>6$ objects GN-z11 \citep{Bunker2023}, RXCJ2248-ID \citep{Topping2024}, and GLASS-z12 \citep{Castellano2024}. 
In fact, every He II emitter in our sample has a He II EW rivaling the one measured for GN-z11 in \cite{Bunker2023}. 
Thus, these diagnostics lend support to the AGN candidacy and demonstrate the remarkable nature of the UV EWs seen in this sample.
\subsection{Broad H$\alpha$}
\label{sec:broad_Ha}
We find evidence of broad H$\alpha$ (FWHM$\gtrsim$1000km/s) for one non-LRD. \cite{Roberts-Borsani2024} also recognized this object (11254, z=6.87) as a broad-line AGN candidate. 
We show the narrow-only and two component fits and their resulting residuals in Figure \ref{fig:11254}. We also include the UV region because of the $>2\sigma$ evidence for a few of the fitted lines.

\subsection{[Ne V]}
\label{sec:neon_detections}
We detect [Ne V] above 5$\sigma$ for LRD 45924 and for narrow-line AGN candidate 45092. None of the 3$\sigma$ limits for the other emitters come within a factor of 10 of the fitted flux for 45092, whereas three upper limits are consistent with the measured flux for 45924. 
Both objects are then consistent with the AGN and composite regions (but not the SFG or Pop III/IMBH ones) of the [O III]/$\mathrm{H\beta}$ vs. [Ne V]/[Ne III] diagnostic \citep{Cleri2023}.
\begin{figure}
    \centering
    \includegraphics[width=\columnwidth]{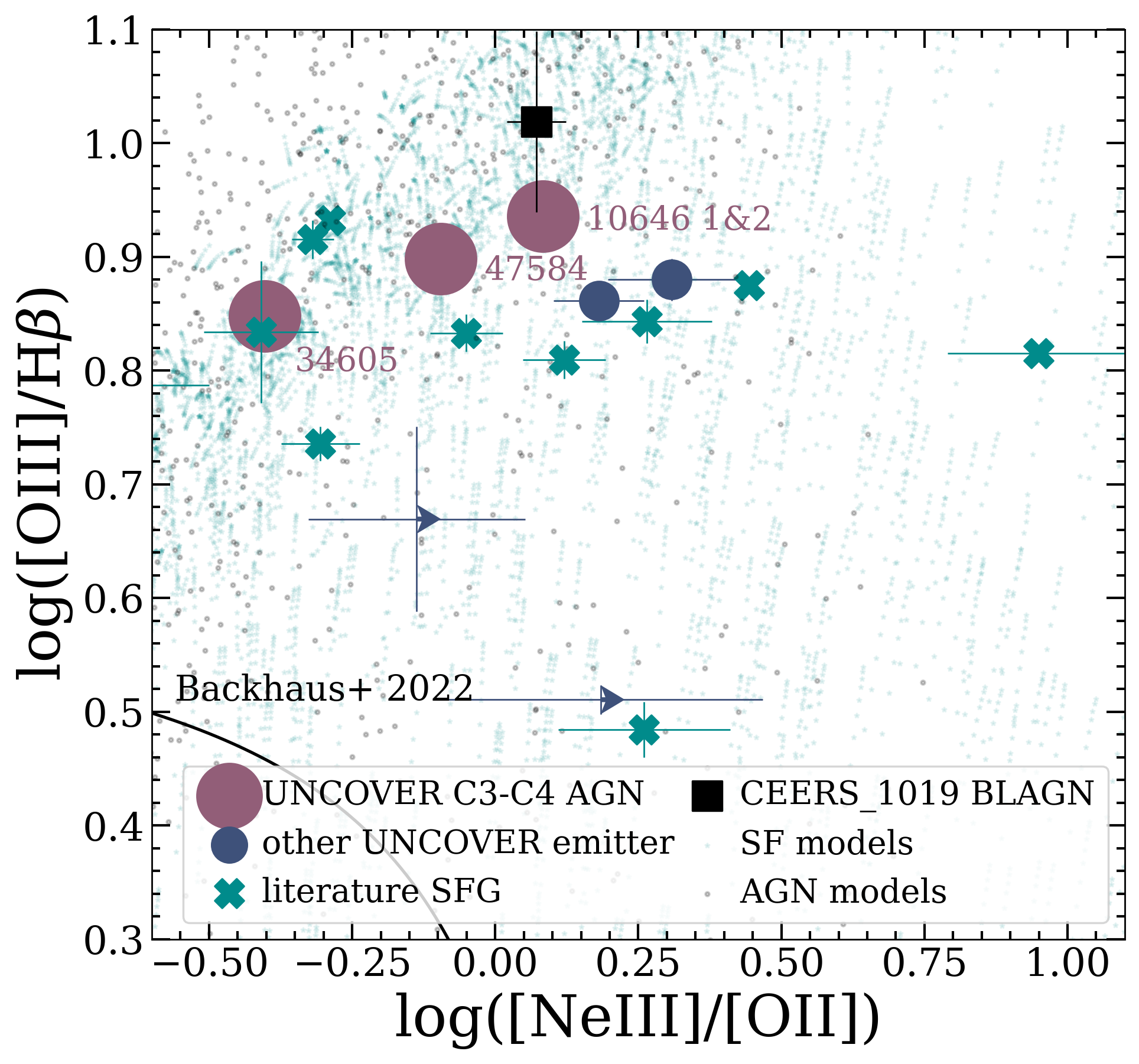}
    \caption{OHNO diagram for literature sources and for our sample of line emitters (circles). 
    Our AGN candidates (Figure \ref{fig:ratio_diagnostics}) are plotted larger than the emitters that are more consistent with the SFG models. Arrows indicate 3$\sigma$ limits for our emitters with either [Ne III] or [O II] undetected.
    C IV emitters 23604 and 36755 are not plotted because of their lack of [Ne III] or [O II] detections. Likewise, AGN candidate 45092 has no [Ne III], [O II], or $H\beta$ detected, with an extreme log([O III]/H$\beta$) lower limit of 2.7. BLAGN 45924 has evidence of [Ne III] and [O II], but the region is too complex for a reasonable fit.
    CEERS\_1019 is a high-z broad-line AGN \citep{Larson2023} and sits at high [O III]/H$\beta$.
    Also plotted are SFGs \citep{Fosbury2003,Christensen2012,Tang2023,Troncoso2014,Trump2023,Vanzella2020} and photoionization models \citep{Feltre2016,Gutkin2016,Hirschmann2019}.}
    \label{fig:OHNO}
\end{figure}
\subsection{OHNO}
\label{sec:OHNO}
In Figure \ref{fig:OHNO} we compare the relative locations of our objects and several literature sources on the OHNO diagram. Consistent with other high-redshift and low-metallicity studies, we see that all of our objects are above the \cite{Backhaus2022} cutoff, implying that they would be selected as AGN in this diagnostic, although the SF photoionization model grid \citep{Gutkin2016,Hirschmann2019} extends throughout the space. 
We thus confirm that this diagram does not reliably separate AGN and SFGs. 
Lastly, while we plot very few literature SFGs, we do see that the star-formation--powered sources do not go to lower [O III]/H$\beta$ than the He II and C III] emitters, so the model grid cut also applies to them.

\subsection{Characterization of Our Line Emitters}
\label{sec:characterization}
These fits and diagnostics leave us with four narrow-line AGN candidates, two broad-line AGN, and seven objects requiring higher-resolution data for classification. 
In particular, we designate the five He II emitters (Figure \ref{fig:he2_spectra}) as AGN candidates. Only one of them sits on the SFG model grids, but that is to be expected considering it is a known LRD with Balmer line widths of $\sim$4500 km/s \citep{Greene2024}. 
To this group we can add 11254, an object with a broad H$\alpha$ component and $\gtrsim 2 \sigma$ evidence of C IV, O III], and C III] emission.

The remaining seven emitters (Figures \ref{fig:c3_spectra} and \ref{fig:c45_spectra}) lack He II detections, precluding strong AGN candidacy with this approach. However, they are all consistent with both the AGN and SFG models in the UV diagnostics.
These C III] and C IV emitters are either AGN or exceptional star-formers with EWs that further challenge models of SFGs.

We provide individual descriptions of the AGN candidates in Appendix \ref{sec:UV_fits}, but note here that of the narrow-line AGN candidates, one (45092) has clear detections of both [Ne V] and X-ray emission, two (45092 and component 2 of 10646) present C IV emission, and two (45092 and 47584) have evidence of N III]. 
Like 45924, the strongest narrow-line AGN candidate (45092) also has N IV]. 
We must consider that candidates without C IV detections, which we consider to be lower-confidence anyway (Section \ref{sec:candidacy}), may be compact, GN-z11-like objects, with metal-poor (but nitrogen-enhanced) bursty star-formation \citep[][]{Bunker2023,Marques-Chaves2023}. 

\section{Galaxy Properties}
\label{sec:properties}
In this section, we comment on the host galaxy properties of the line emitters, both candidate AGN and exceptional star-formers. 
The opportunity to derive these properties is another benefit of the extensive wavelength coverage at our disposal with JWST. 
In the presence of a narrow-line AGN, it is reasonable to assume that the continuum is galaxy-dominated \citep[e.g.,][]{Hickox2018}, but higher-resolution data could determine the AGN contribution to the photometry.
We present these host galaxy properties in three figures and, in each case, compare the emitter sample(s) to the parent sample of $z>3$ UNCOVER galaxies with spectroscopic follow-up. We note that the parent sample itself involves a complex selection function \citep{Price2024}. As such, to span the range of specific star formation rate (sSFR) and stellar mass, we rely on the UNCOVER \textit{photometric} sample of galaxies with $\mathrm{z_{phot}>3}$. Wherever available, these photometric redshifts and properties incorporate the full medium band coverage provided by Cycle 2 program ``Medium Bands, Mega Science" \citep{Suess2024}. In Table \ref{tab:props}, we list the $\mathrm{16^{th}}$, $\mathrm{50^{th}}$, and $\mathrm{84^{th}}$ percentiles of these properties for the parent sample, emitter sample, and emitter subsamples.

Here, we explain the three figures and overall differences between the parent sample and the sample of \emitternum~emitters. In Sections \ref{sec:gp_He2}--\ref{sec:gp_C4}, we then highlight commonalities within each subsample of line emitters. First, we mark the final sample of emitters in magnitude vs. redshift space and include the \cite{Greene2024} LRDs to compare to the locations of known broad-line AGN (Figure \ref{fig:mag_redshift}). The line emitters span the parent sample range in both axes but are systematically brighter. 

In Figure \ref{fig:ssfr-mass}, we plot sSFR averaged over 100 Myr vs. stellar mass for the sample of line emitters, with contours for the photometric and spectroscopic $z>3$ samples. 
Almost all the emitters are at high mass and/or high sSFR relative to the parent sample, with none within the one $\sigma$ contour of the $\mathrm{z_{phot}}>3$ sample. 
In particular, the AGN candidates are clearly biased to high stellar masses, while the other emitters are lower mass than the AGN candidates but present modestly higher sSFRs than typical in the parent sample.

\begin{figure} 
    \centering
    \includegraphics[width=\columnwidth]{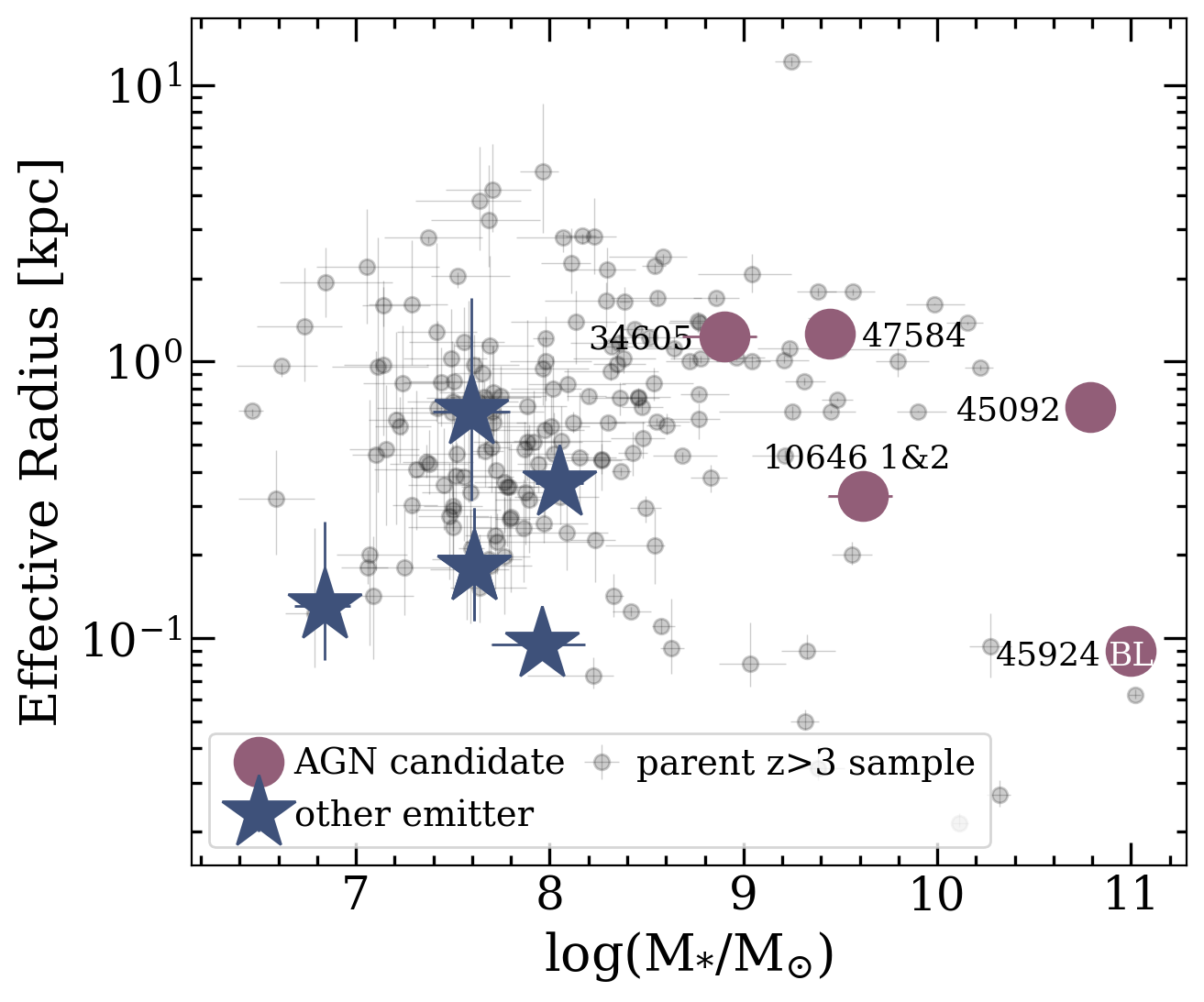}
    \caption{Magnification-corrected effective radius in the F444W band vs. stellar mass for line emitters with reliable size measurements and for the parent $z>3$ sample. The AGN candidates, which are the He II emitters, are plotted with red circles and labeled with their MSA IDs, while the other emitters are marked with blue stars. 
    The two components of 10646 have a single size measurement, and thus this value serves as an upper limit for each component.
    The C III] emitter 17467 is significantly sheared and thus excluded here.
    The stellar population of LRD 45924 is difficult to model, partially because of the spectral resolution and the luminosity of the AGN component. The plotted stellar mass of $10^{11} \mathrm{M_{\odot}}$ comes from the preferred fit. Other potential solutions all require $\gtrsim10^{10} \mathrm{M_{\odot}}$ \citep{Labbe2024}.}
    \label{fig:size-mass}
\end{figure}

Lastly, we compare stellar masses to effective radii in Figure \ref{fig:size-mass} and compare to the spectroscopic parent sample. A handful of emitters are remarkably compact, especially for their mass, resulting in statistical significance to the size difference for the emitters and the parent sample. We now discuss the locations of the subsamples of emitters with He II, C III], and C IV detected. 

\subsection{AGN Candidates (He II Emitters)}
\label{sec:gp_He2}
The He II emitters are the clearest AGN candidates in the ratio-ratio and EW-ratio diagnostics (Figures \ref{fig:ratio_diagnostics} and \ref{fig:EW_diagnostics}), or, in the case of 45924, known broad-line AGN. These objects are all relatively bright and massive, with these differences being the most significant among the comparisons in this section.

The He II emitters are all brighter than $\mathrm{25^{th}}$ mag in F444W.
Relatedly, they are more massive than the other emitters and more massive than is typical in the parent spectroscopic and photometric samples. 
The strongest narrow-line AGN candidate (45092) has the highest mass in the parent sample and sits at low sSFR, as it is a post-starburst galaxy (Bezanson et al. in prep). The closest parent sample object to 45092 in sSFR-mass space is 18407, a massive ($\mathrm{log(M_{*}/M_{\odot})}\sim 10.3$) quiescent galaxy at $z=3.97$ \citep[][]{Setton2024}.
On the other hand, a couple of the other exceptionally high-mass He II emitters are at high sSFR. 

Because this group is so massive, there are few comparable galaxies for size comparison. Still, they seem compact for their masses, although that observation is not statistically significant. The least compact AGN candidates are 34605 and 47584, which also lack C IV detections. 
On the other hand, for either F444W magnitude or mass, a two-sample Kolmogorov-Smirnov (KS) test comparing the He II emitters with the parent sample shows statistical significance (p-value $<$ 0.05). 
This result implies that the He II mass and magnitude distributions are different from those of the parent sample; this conclusion is also clear in Table \ref{tab:props}.
We further address this mass bias in Section \ref{sec:discussion}.

\subsection{C III] Emitters}
\label{sec:gp_C3}
Of the subsamples, the C III] emitters have the closest distributions to the parent sample, particularly if we exclude the four that are also He II emitters.

\begin{figure}
    \centering
    \includegraphics[width=\columnwidth]{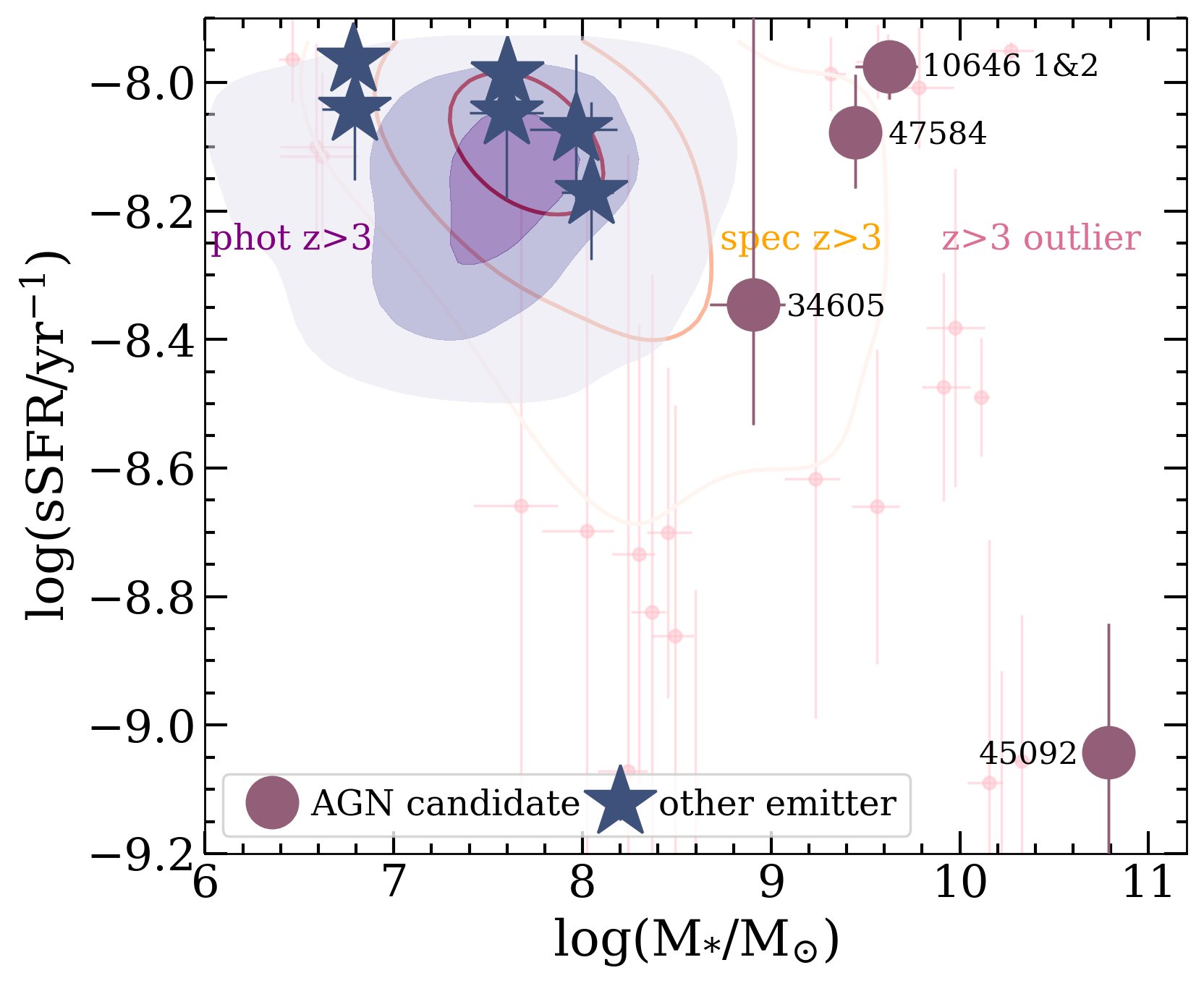}
    \caption{100 Myr specific star formation rate vs. stellar mass for the line emitters, with contours for the UNCOVER photometric and spectroscopic $z>3$ samples shown for reference. We also plot individual parent sample outliers in pink. 
    The AGN candidates, which are the He II emitters, are plotted with red circles and labeled with their MSA IDs, while the other emitters are marked with navy stars. 
    The two components of 10646 share a single sSFR estimate, so we combine them here. In Table \ref{tab:props}, we use their individual log($\mathrm{M_{*}/M_{\odot}}$) estimates of 8.93 and 9.06 for components 1 and 2, respectively.
    We exclude LRD 45924 because of the significant uncertainty on its sSFR \citep{Labbe2024}.}
    \label{fig:ssfr-mass}
\end{figure}

The C III] emitters are also under-represented in number at the faintest magnitudes, but their magnitudes do range from 21--28.  
As is apparent in Figure \ref{fig:ssfr-mass} or a KS test, 
there is not sufficient evidence to conclude that these emitters have different sSFR or mass distributions than the parent sample. However, if we exclude the He II emitters from this calculation, the remaining C III] emitters do have significantly higher sSFR.
C III] emitter 17467 is the galaxy with the highest sSFR in the sample. We also note that it is magnified by a factor of 133, or 25 times more than the next most magnified source in the emitter sample.

C III] emitters are, on average, more compact compared the parent sample, although a KS test does not allow us to reject that their sizes are drawn from the same distribution. Still, broad-line AGN 45924 is more compact than the other line emitters, and this mean difference holds up whether or not we include it in the average for this subsample.

\subsection{C IV Emitters}
\label{sec:gp_C4}
The C IV emitters are the most compact subsample; the five most compact emitters are all C IV emitters.
They are also potentially over-represented at the highest redshifts: three of the seven C IV emitters are above redshift 6, while this is only true for 22\% of the parent sample. Furthermore, in the absence of the AGN candidates, they are typical in mass and high in sSFR. 
Although the C IV emitters do not generally skew to higher masses, the two most massive emitters are precisely the He II emitters with C IV also detected. 

\begin{table*}
    \centering
    \begin{tabular}{cccccc}
        \hline
         & Parent (\parentnum) & Emitters (\emitternum) & He II (5) & C III] (8) & C IV (7) \\
        \hline
        \hline
        F444W mag &$27.4^{+1.09}_{-1.98}$&$24.85^{+3.18}_{-1.41}$&$23.83^{+0.44}_{-2.06}$&$24.58^{+2.97}_{-0.57}$&$27.87^{+0.64}_{-5.72}$\\ 
        z &$4.47^{+1.89}_{-0.99}$&$4.47^{+3.56}_{-0.84}$&$3.97^{+2.61}_{-0.61}$&$3.99^{+1.94}_{-0.17}$&$5.96^{+2.05}_{-2.15}$\\
        $\mathrm{log(M_{*}/M_{\odot}})$ &$8.02^{+1.02}_{-0.53}$&$8.65^{+1.57}_{-1.16}$&$9.30^{+1.23}_{-0.31}$&$8.68^{+0.49}_{-1.15}$&$7.82^{+2.30}_{-0.30}$\\
        log(100 Myr sSFR / $\mathrm{y^{-1}}$) &$-8.14^{+0.11}_{-0.32}$&$-8.06^{+0.08}_{-0.20}$&$-8.19^{+0.17}_{-0.39}$&$-8.03^{+0.06}_{-0.22}$&$-8.04^{+0.06}_{-0.11}$ \\
        $\mathrm{r_{eff}}$ [kpc] &$0.66^{+0.73}_{-0.43}$ &$0.34^{+0.64}_{-0.23}$&$0.96^{+0.29}_{-0.46}$&$0.34^{+0.49}_{-0.09}$&$0.25^{+0.41}_{-0.13}$\\
        log([O III]/$\mathrm{H\beta}$)&$0.68^{+0.19}_{-0.22}$&$0.84^{+0.05}_{-0.31}$&$0.87^{+0.04}_{-0.03}$&$0.87^{+0.04}_{-0.10}$&$0.61^{+0.24}_{-0.11}$\\
        \hline
    \end{tabular}
    \caption{Galaxy property medians for the parent sample, line emitter sample, and the emitter subsamples. log([O III]/$\mathrm{H\beta}$) values only include sources with both lines detected, and thus exclude AGN candidate 45092, which has no detected $\mathrm{H\beta}$. 
    The components of 10646 are separate for magnitude and mass (Weaver et al. in prep) but combined for sSFR, log([O III]/$\mathrm{H\beta}$), and effective radius.
    The only $\mathrm{16^{th}-84^{th}}$ percentile range inconsistent with the parent sample's are the F444W magnitudes of the He II emitters.}
    \label{tab:props}
\end{table*}

\section{Discussion and Summary}
\label{sec:discussion}
We have demonstrated the usefulness of NIRSpec/PRISM for the identification of high-redshift (e.g., $z\gtrsim3$) narrow-line AGN candidates.
Central to these efforts are careful deblending and the use of the sample's [O III]/H$\beta$ to cut down on the relevant SFG models. The role of this cut underscores the essential constraining power of the PRISM wavelength range; no JWST medium-resolution gratings could capture both sets of lines in one integration.

We select \emitternum~strong C IV$\lambda\lambda$1548,1551, He II$\lambda$1640, and CIII]$\lambda\lambda$1907,1909
emitters from a sample of \parentnum~$z>3$ galaxies. 
One He II emitter (45924) is a known LRD while the other four are the most promising narrow-line AGN candidates. 
We identify one additional broad-line AGN (11254, Figure \ref{fig:11254}).
If the remaining seven emitters are indeed SFGs, their UV EWs largely require updated models. 

Here, we discuss the implications of the line emitter properties, contextualize the AGN candidates using other JWST studies, and highlight opportunities to expand on this work through higher-resolution data. 
\subsection{NLAGN Candidates}
The He II emitters, which include the best narrow-line AGN candidates, are brighter and higher-mass than the other emitters and are outliers in the parent sample; 17\% of galaxies in the parent sample with $\mathrm{log(M_{*}/M_{\odot})}>2\times10^9$ are also AGN candidates. 
If we assume we are unbiased for the highest-mass hosts, this result can be compared to the \cite{Maiolino2023} finding that 10\% of $4<z<6$ galaxies host broad-line AGN with $\mathrm{L_{AGN}>10^{44} \, erg/s}$.

The other galaxies in the line emitter sample have 3$\sigma$ line ratio limits consistent with both model grids.
As a result, the AGN candidate separation could be a selection effect in which we are insufficiently sensitive to detect He II in lower-mass objects, because they likely host lower-mass SMBHs, which would have weaker emission lines if they are otherwise comparable.

Comparing to other efforts focused on narrow-line AGN candidates, our findings are not inconsistent with rates found in JADES \citep[][]{Scholtz2023}, although we are hesitant to quantify an exact AGN fraction because of our spectroscopic selection function.
In particular, \cite{Scholtz2023} presented 27 NLAGN candidates at $z>3$. Most candidates required a conservative version of optical diagnostic diagrams for selection and higher-resolution spectra yielded the six candidates identified with C IV, C III], He II. 
The clearest difference is the candidate mass range: none of their $z>3$ candidates have $\mathrm{log(M_{*}/M_{\odot}})>9$, further supporting the possibility that we can only detect the highest-mass AGN with PRISM and rest-frame UV diagnostics. Likewise, without a well-defined selection, it is difficult to conclude anything about redshift evolution.

\subsection{C IV emitters}
The C IV emitters seem to include AGN and likely star-formers, but trace a potentially different population from the C III] emitters. Three C IV emitters are He II emitters and AGN candidates, while the other four are consistent with either population but all at high sSFR.
Key properties potentially setting C IV emitters apart are their compactness, their relatively high redshift distribution, and their lower [O III]/H$\beta$.
However, with the exception of compactness, this conclusion is not robust with the current sample size (Table \ref{tab:props}).
Additionally, 36755 and 23604 boast comparable C IV EWs ($88^{+18}_{-15}\mathrm{\mathring{A}}$ and $51^{+14}_{-14}\mathrm{\mathring{A}}$, respectively) to the exceptional $z>6$ galaxies investigated in \cite{Topping2024} and \cite{Castellano2024}.

C IV emission is seen in a number of contexts: AGN (Figure \ref{fig:literature_diagnostic}), $z>5$ low-metallicity sources \citep{Stark2015,Mainali2017,Schmidt2017,Witstok2021}, and Lyman continuum emitting galaxies \citep[e.g.,][]{Schaerer2022}. It is unclear the extent to which properties like metallicity, SFR, ionization parameter, and ISM density play a role in the presence and strength of C IV emission.
In the case of C IV-emitting star-forming galaxies, metal-poor gas and young stellar populations seem essential \citep{Senchyna2017,Topping2024CN}; these requirements may explain the high redshift and relatively low [O III]/H$\beta$ of our C IV emitters.
However, the JADES $z<6$ Ly$\alpha$ emitter stack has higher C IV EW ($39\pm3\mathrm{\mathring{A}}$) than the higher-redshift one \citep[$9\pm3\mathrm{\mathring{A}}$,][]{Kumari2024}, and real redshift evolution is difficult to establish definitively given the changing spectral resolution.

Lastly, we note an apparent connection to Ly$\alpha$ that is consistent with literature findings. 
C IV emission can be a tracer for the escape of Lyman continuum photons \citep[e.g.,][]{Naidu2022,Mascia2023} and among $z>3$ faint Ly$\alpha$ emitters, \cite{Feltre2020} only detected C IV in emission in their stack with the highest Ly$\alpha$ EWs (i.e., $\gtrsim100\mathrm{\mathring{A}}$). 
This discrepancy is likewise apparent in the median spectra of LAEs and non-LAEs across public NIRSpec PRISM data, despite the stacks having the same average redshift \citep{Roberts-Borsani2024}.
In our sample of line emitters, high EW Ly$\alpha$ emission is common among the C IV emitters but not in the other subsamples. However, detailed quantification of this difference is beyond the scope of this work. 

\subsection{The Future of NLAGN with JWST}
In the first couple years of JWST, AGN identification has focused on broad-line objects \citep{Harikane2023,Kocevski2023,Larson2023,Maiolino2023,Ubler2023}. Considering the surprisingly high number densities \citep[e.g.,][]{Greene2024}, we could be missing an even larger obscured population at high redshifts.
We have demonstrated both the opportunities and obstacles for identifying NLAGN using JWST.
Updated models are necessary because of the unprecedented EWs presented here, but there is also more to do with existing models.
Although we illustrated the usefulness of the wavelength range, we have not compared AGN vs. SFG model fits to the full PRISM spectra.
Additional constraints will come from the continuum shapes and numerous other detectable emission lines unexplored in this work. 

By modeling the full observed wavelength range, one can also isolate which features we cannot explain with self-consistent modeling \citep[e.g.,][]{Cameron2024,Tacchella2024,Vidal-Garcia2024}.
As we discuss in Section \ref{sec:candidacy}, current photoionization models fall short for known SFGs, especially in modeling He II, making it hard to know the reliability of AGN diagnostics that use this line.
Increasingly clear is the necessity of including stellar models with non-solar abundance ratios \citep[e.g.,][]{Grasha2021}. New binary models may reconcile these discrepancies with observations \citep[e.g.,][]{Lecroq2024}.
Another promising approach is the use of a flexible, source-agnostic tool that can accurately reconstruct the ionizing spectrum and whether binary stellar models are necessary for its production \cite[\texttt{Cue},][]{Li2024}.

Larger samples and higher spectral resolution would make the use of higher-ionization lines more feasible. At the same time, the tension with models will be more apparent, particularly at low metallicity and high ionization parameter.
Although this approach is necessary, the role of PRISM in these efforts will persist, because of its ability to efficiently observe a broad wavelength range. 
For instance, a 7$\sigma$ [Ne V] detection in a z=5.59 galaxy required NIRSpec high-resolution and an integration time of 14.7 hours \citep{Chisholm2024}. The [Ne V] flux was 24 times less than that of H$\alpha$, with a rest-frame EW of $\sim$11$\mathrm{\mathring{A}}$. The EW and strength relative to other emission lines are significantly higher than those found in star-forming galaxies with [Ne V] detections \citep[e.g.,][]{Izotov2004,Izotov2012,Izotov2021}, supporting the case for the line's AGN origin.
Only four of our objects have 3$\sigma$ upper limits ruling out that they have at least the same [Ne V] flux as this object.
Only one object in the \cite{Scholtz2023} JADES sample boasted a [Ne V] detection and He II and C III] were not detected in this object, making it difficult to pinpoint relative to AGN and SFG models. 

We encourage additional multi-wavelength and high-resolution follow-up of the AGN candidates and exceptional star-formers presented here; among the potential SFGs, we emphasize MSA IDs 37005, 36755, and 23604 because of their unprecedented UV EWs and 17467 because of its low mass, high sSFR, and the extent of its magnification. There is also abundant PRISM data that is unexplored with this approach \citep{Roberts-Borsani2024}. Although the PRISM rest-frame UV selection method presented here comes with myriad challenges, it has the potential to uncover poorly-understood high-redshift NLAGN and metal-poor populations, especially with the help of modeling advancements and higher-resolution follow-up.

\acknowledgements
We thank Anna Feltre and Michaela Hirschmann for sharing their models and for helpful discussions.

HT and JS acknowledge support by the National Science Foundation Graduate Research Fellowship Program under Grant DGE-2039656. 
Any opinions, findings, and conclusions or recommendations expressed in this material are those of the authors and do not necessarily reflect the views of the National Science Foundation.

TBM was supported by a CIERA Postdoctoral Fellowship.

LJF acknowledges support by grant No.~2020750 from the United States-Israel Binational Science Foundation (BSF) and grant No.~2109066 from the United States National Science Foundation (NSF), by the Israel Science Foundation Grant No.~864/23, and by the Ministry of Science \& Technology, Israel.

PD acknowledges support from the NWO grant 016.VIDI.189.162 (``ODIN") and warmly thanks the European Commission's and University of Groningen's CO-FUND Rosalind Franklin program. 

Support for this work was provided by The Brinson Foundation through a Brinson Prize Fellowship grant. 

DM and RP acknowledge funding from JWST-GO-02561-013 and JWST-GO-04111.035, provided through a grant from the STScI under NASA contract NAS5-03127.

This work is based in part on observations made with the NASA/ESA/CSA \textit{James Webb Space Telescope}. The data were obtained from the Mikulski Archive for Space Telescopes at the Space Telescope Science Institute, which is operated by the Association of Universities for Research in Astronomy, Inc., under NASA contract NAS 5-03127 for JWST. These observations are associated with JWST Cycle 1 GO program \#2561 and Cycle 3 GO program \#4111. The JWST data presented in this article were obtained from the Mikulski Archive for Space Telescopes (MAST) at the Space Telescope Science Institute. The specific observations analyzed can be accessed via \dataset[DOI]{https://doi.org/10.17909/qgpt-c913}. Support for program JWST-GO-2561 was provided by NASA through a grant from the Space Telescope Science Institute, which is operated by the Associations of Universities for Research in Astronomy, Incorporated, under NASA contract NAS5-26555.

\bibliographystyle{aasjournal}
\bibliography{general.bib}

\appendix
\restartappendixnumbering
\section{Fitting Simulated Rest-Frame UV Spectra}
\label{sec:simulations}
To evaluate the efficacy and limits of the MultiNest fitting of the UV lines (Section \ref{sec:fitting}), we analyze mock versions of our final line emitters.
Here, we describe the set-up for these simulated spectra and the key take-aways that inform our analysis of the MultiNest fits to the emitters. In short, MultiNest can recover injected flux values and provide reliable uncertainties.
Furthermore, we confirm that the presence of correlated noise affects these mock results, increasing the chance of false detections at low SNR. However, these modest contributions do not significantly change our interpretation of the results.
Although our fitting approach can deblend lines sufficiently well for our purposes, 
blended line ratios can be more biased than individual fluxes, since a slight underestimation of one line frequently corresponds to an overestimation of the other component of the blend. However, such a bias may be restricted to ratios that include a non-detection.
We also highlight the few line emitters that could be false positives, albeit with low probability. 

As the AGN candidates, the He II emitters constitute the most important group in this work. They also present the most difficult deblending problem since He II is included in all the diagnostics. C III] is blended with Si III], but C III] should be stronger than Si III] and the strength of either of those lines relative to He II bolsters the case for star formation as the origin of the ionization.
We therefore perform MultiNest fits to many mocks of each He II emitter to check our confidence in the He II detections. This approach addresses a few potential concerns, with the key one being a check of the likelihood that we are attributing significant He II flux to what is really O III]. 

First, for each He II emitter, we use MultiNest to fit 100 mock spectra with parameters equal to the best fit ones for the real spectrum. For each iteration, we add Gaussian noise to each spectral bin. For the standard deviation of this noise, we use the true error array scaled by the best-fit multiplicative error scaling. In the case of 45092, 45924, and 47584, we use the original error array and mirror the original fits by not including an error scale factor in the model of the mock spectra. 

\begin{figure}
    \centering
    \includegraphics[width=\columnwidth]{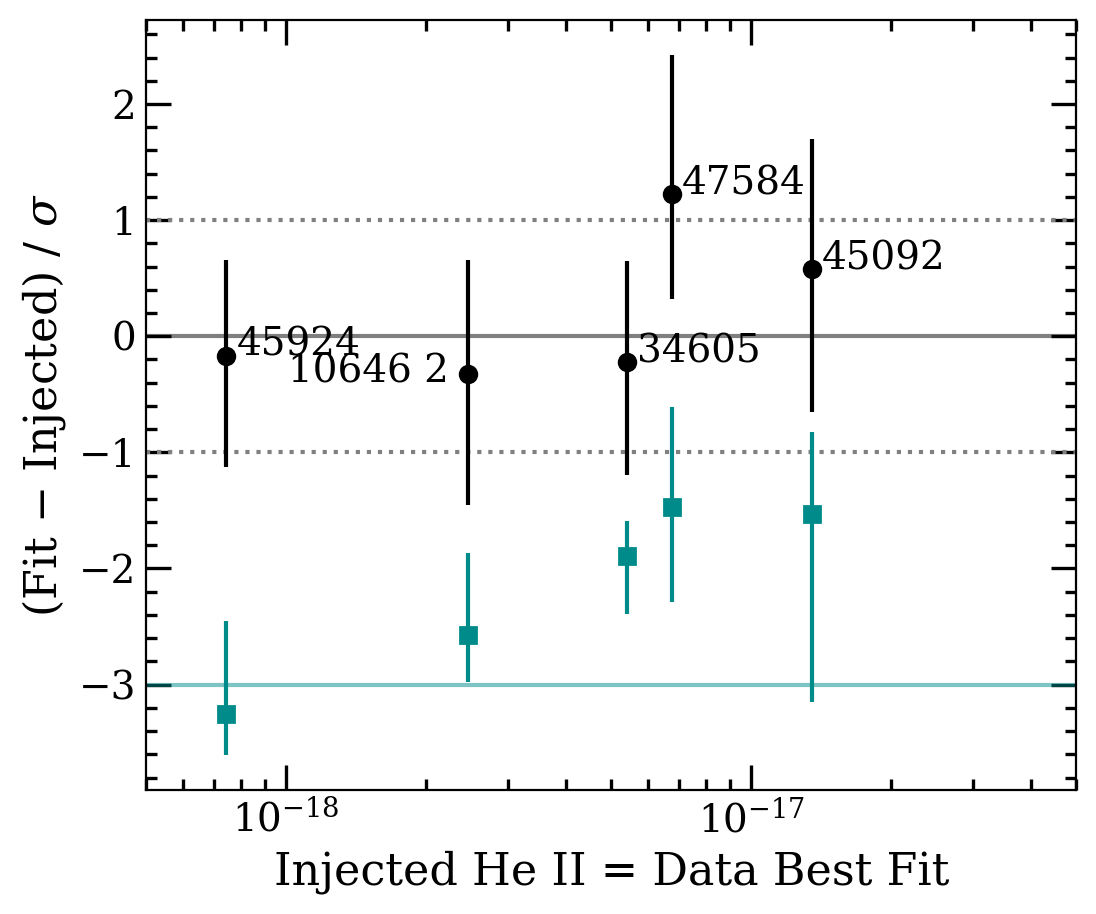}
    \caption{Summary of fits to mock versions of He II emitters. The black points and error bars use the $\mathrm{16^{th}}$, $\mathrm{50^{th}}$, and $\mathrm{84^{th}}$ percentiles
    of the He II fits to the data's best fit model plus uncorrelated noise. Because the injected value is subtracted off, points should be centered at zero. These differences are then scaled by the He II one $\sigma$ error from the best fit to the data, such that an error bar length of one indicates perfect agreement between the original MultiNest error bar and the error based on the mocks.
    Also plotted, in cyan squares, are the results from injecting He II at the 3$\sigma$ lower limit from the original fit to the data. We still use the best fit values to place these points on the y-axis, such that these squares would ideally sit at $-$3.
    }
    \label{fig:mocks}
\end{figure}

We then repeat this test but with the injected He II equal to the 3$\sigma$ lower limit from the original fit to the data and the O III] adjusted such that the total blend flux remains the same. Through this set-up, we confirm that these kind of underlying flux values do not yield the best-fit He II values that we found with the real data. In other words, we do not need to be concerned that our He II detections are actually O III] or noise.
Figure \ref{fig:mocks} summarizes the findings from these two tests. Namely, for our purposes, MultiNest successfully identifies the He II flux values and the associated uncertainties. 
Moreover, none of the fits to the He II 3$\sigma$ lower limits overlap with the best fit He II values, which we interpret as indication that we have not falsely attributed He II flux to O III] or noise.

A slight modification on this approach also allows us to understand the effect of correlated noise, which is surely present but has not yet been robustly quantified in UNCOVER. Once available, one should work with a covariance matrix when fitting (as in JADES et al. in prep).
In particular, we create mocks with the injected He II at the 3$\sigma$ lower limit but construct the added noise to have correlations between neighboring pixels.
We perform this test with correlation coefficients of 0.3 and 0.6, finding that only low SNR cases (i.e., 34605) present a clear trend with increasing correlation coefficient. Namely, while 1\% of uncorrelated noise mocks yield fitted fluxes above the real spectrum's best-fit He II value, this rate increases to 2\% and 4\% for correlation coefficients of 0.3 and 0.6, respectively. Regardless of the specifics of the noise, it is clear that He II emitter 34605 has some chance of being a false positive.

\begin{figure*}
    \centering
    \includegraphics[width=\columnwidth]{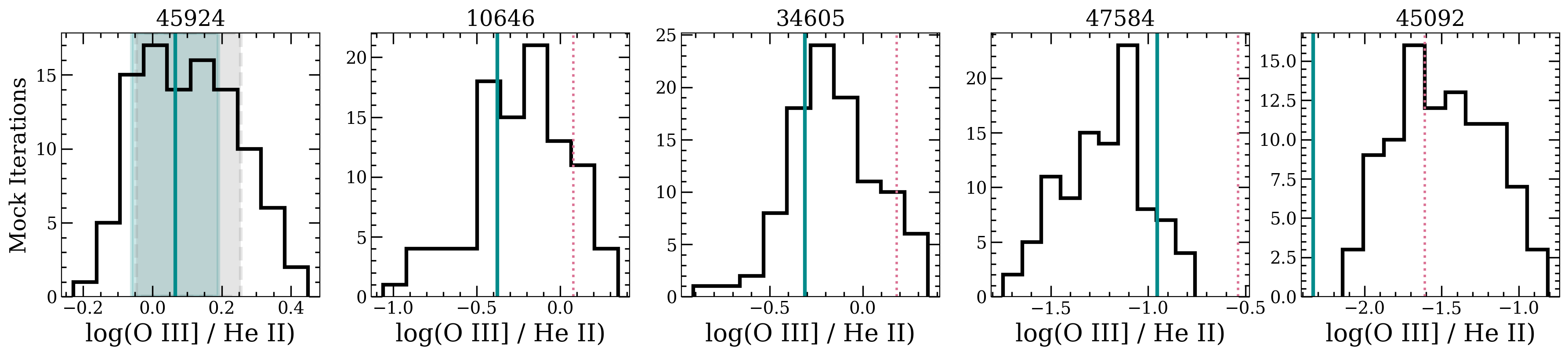}
    \caption{Distribution of fitted log(O III]/He II) across 100 mock iterations for each He II emitter. The cyan vertical lines mark the injected ratio, highlighting the bias for 45092 and 47584.
    For 45924, the only one with an O III] detection, we include cyan shading for the one sigma uncertainty from the fit to the real data and grey shading for the one sigma uncertainty based on the plotted Multinest mock iterations.
    The remaining objects' histograms include a pink dotted line for the original fit's 3$\sigma$ upper limit, which we use in the diagnostics.
    }
    \label{fig:ratio_hists}
\end{figure*}

We also examine potential bias in the ratio of O III]/He II, which is one of the key diagnostic ratios. Here, we have more cause for concern, since the uncertainties in each line are correlated, leading to systematic bias in the ratio.
First, our 3$\sigma$ O III]/He II upper limits for sources without O III] detections may be unnecessarily high, since the fits attribute too much flux to the undetected O III] line.
In particular, as shown in Figure \ref{fig:ratio_hists}, for the source with the lowest injected O III] flux and O III]/He II (45092), the mock fits are biased towards higher O III]/He II ratios. 
The median ratios from the mocks are just outside the 1$\sigma$ ratio uncertainties from the best fit to the data, but we indicate these ratios as upper limits in the diagnostics, anyway. This bias simply indicates that the true O III]/He II ratios are likely lower than the upper limits indicate, making the sources even clearer AGN candidates.

On the other hand, the O III]/HeII ratios of 47584 are under-estimated.
Thus, the O III] flux may drive the direction of the bias, since 47584 had the highest injected O III] flux, as well as the most over-estimated He II flux (Figure \ref{fig:mocks}).
This bias is slight, though, and such an offset would be inconsequential on the diagnostics.
However, for the one He II emitter with O III] also detected (45924), the MultiNest and mock O III]/He II ratios are in good agreement.

These mocks also allow us to test the redshift fitting in this region. The redshift fits work well, with a mean deviation from the injected value of $\lesssim$0.002. There are rarely any fits close to the 0.01 prior bounds. Furthermore, the success of the redshift fitting does not rely on a strong C IV flux, despite C IV being one of the two unblended lines and the only one to sometimes be strong in the real spectra.

Lastly, for all \emitternum~line emitters, we perform an additional set of fits using the error arrays to inject noise, but no emission line flux, to evaluate how often we recover a false positive significant line.
We note that there is a false positive $\sim$1\% of the time for 34605, while the others do not show false ``detections" across 300 mocks with uncorrelated noise. However, $\sim$2\% of the mocks for 23604 and 10155 yield $3-4\sigma$ C IV measurements, further supporting our use of a $5\sigma$ cutoff for C IV.

\section{Rest-Frame UV Fits to Line Emitters}
\label{sec:UV_fits}
Here, we present the emitters with their MultiNest fixed- and free-redshift fits. We also include some details about each of the AGN candidates.

\subsubsection{10646 2}
10646 (z=8.51) seems to be two objects, resolved in the UV (Weaver et al., in prep). 
One component, which we denote 10646 2, is a C IV, He II, and C III] emitter, while the other is included only as a C III] emitter, although it has C IV at the 3$\sigma$ level and evidence for He II+O III].
Neither one has compelling evidence for N IV] or N III]. There is a hint of [Ne V] in the combined spectrum, but this feature is also consistent with noise. All in all, 10646 2 is one of the strongest AGN candidates and also the highest-redshift among both the He II and C III] emitters in the sample. \cite{Fujimoto2023bubble} mentioned this object as a potential AGN with a nearby broad-line AGN (MSA ID 20466), both of which are seemingly surrounded by an ionization bubble, as evidenced by emission blueward of Ly$\alpha$.

\subsubsection{45092}
45092 (z=3.46, Bezanson et al. in prep) is a strong AGN candidate with C IV, He II+O III], Si III]+C III], and N IV]. Crucially, 45092 has the clearest, strongest [Ne V] of any source in the parent sample and is detected by \emph{Chandra} (Bezanson et al. in prep). It is also one of the most massive galaxies in UNCOVER with $\mathrm{z_{phot}>3}$.

\subsubsection{47584}
47584 appears to have He II(+O III]), N III], and (Si III]+)C III] but no C IV.  
Because of the N III], the lack of C IV, and the fact that 47584 sits at high sSFR (Figure \ref{fig:ssfr-mass}), we must consider that this object is actually an exceptional low-metallicity nitrogen-enhanced SFG.  

\subsubsection{34605}
We include 34605 because of its He II detection, but it is a questionable candidate because of the marginal detection (Section \ref{sec:simulations}), the lack of strong evidence for any other lines, and the overall noisiness of the spectrum. 

\subsubsection{BLAGN}
LRD 45924 has He II emission (along with a host of other lines) but its line ratios place it in the star-forming region of the UV ratio diagnostics, as is expected for broad lines (Figure \ref{fig:literature_diagnostic}). As mentioned in \cite{Greene2024}, 45924 is the one LRD with decent evidence of [Ne V].

\begin{figure}
    \centering
    \includegraphics[width=0.85\columnwidth]{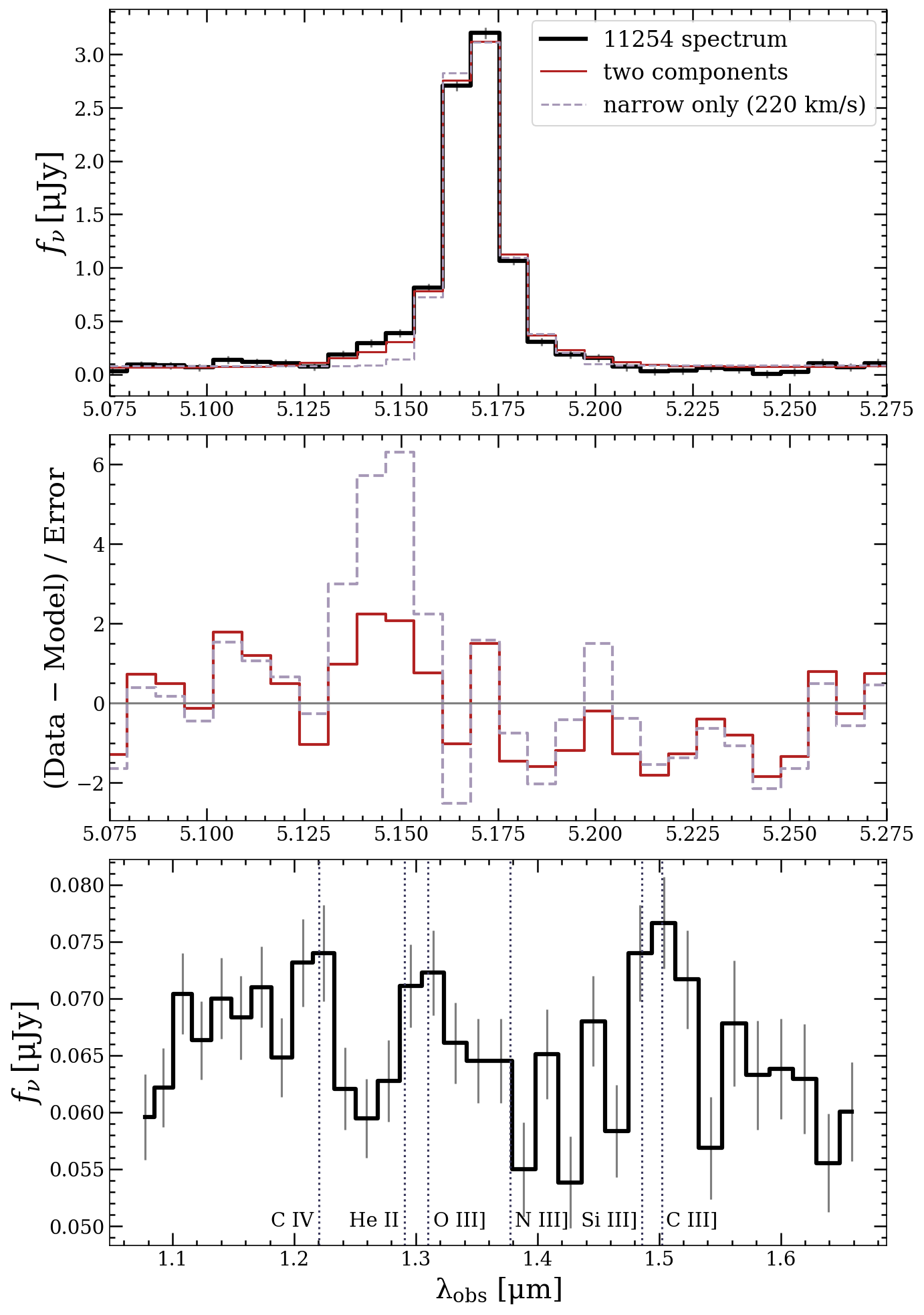}
    \caption{Fits to H$\alpha$+[N II] for 11254 (top) with resulting residuals (middle) and potential UV lines (bottom). The narrow-only best fit (dashed) uses a FWHM of 220 km/s and leaves clear residuals in the wings. The two-component fit includes $\sim$70\% of the total H$\alpha$ flux in a 77 km/s narrow component and the rest in a 2600 km/s broad one.}
    \label{fig:11254}
\end{figure}

We also find 11254 (z=6.87) has a broad H$\alpha$ component, making it another strong AGN candidate. The residuals from a narrow-only fit illustrate the need for this component (Figure \ref{fig:11254}). It also joins 45924 in being more compact than all the other line emitters. 

\begin{figure*}
    \centering
    \includegraphics[width=\columnwidth]{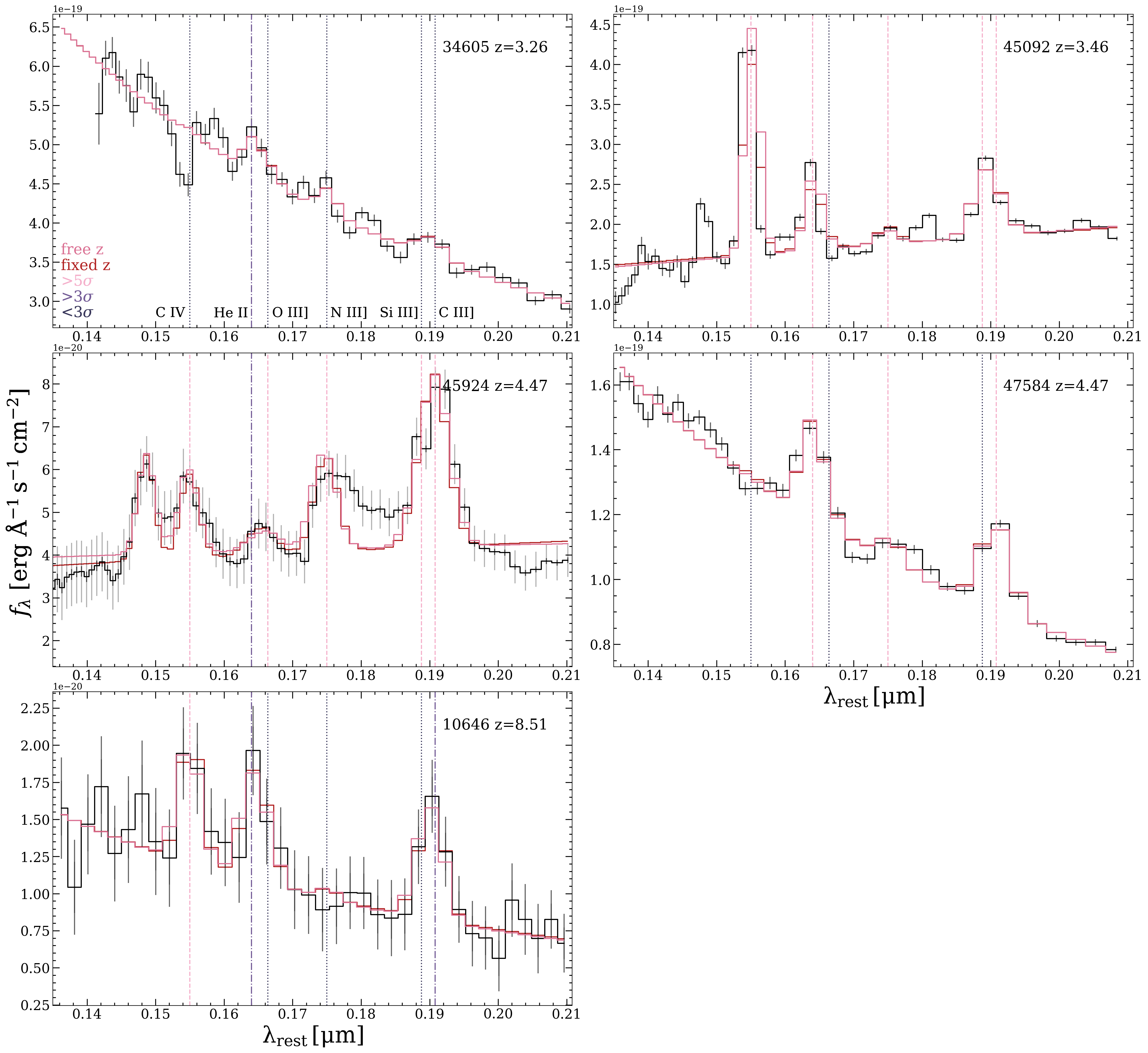}
    \caption{MultiNest fits to C IV through C III] for He II emitters in our sample. The fixed-redshift and free-redshift fits are in red and pink, respectively. The error bars using the best-fit error scaling factor are shown in a lighter grey than the original error bars and the free-redshift fits to 45092, 45924, and 47584 do not use error scaling. Significances of $>5\sigma$, 3$-$5$\sigma$ and $<3\sigma$ are marked with dashed pink, dash-dotted purple, and dotted navy lines. All the He II emitters show an unmodeled blend redward of N III]; this blend can be well-fit by Fe II, [Ne III]+Si II, and Al III, which are all present in the \cite{Shen2019} AGN composite.}
    \label{fig:he2_spectra}
\end{figure*}

\begin{figure*}
    \centering
    \includegraphics[width=\columnwidth]{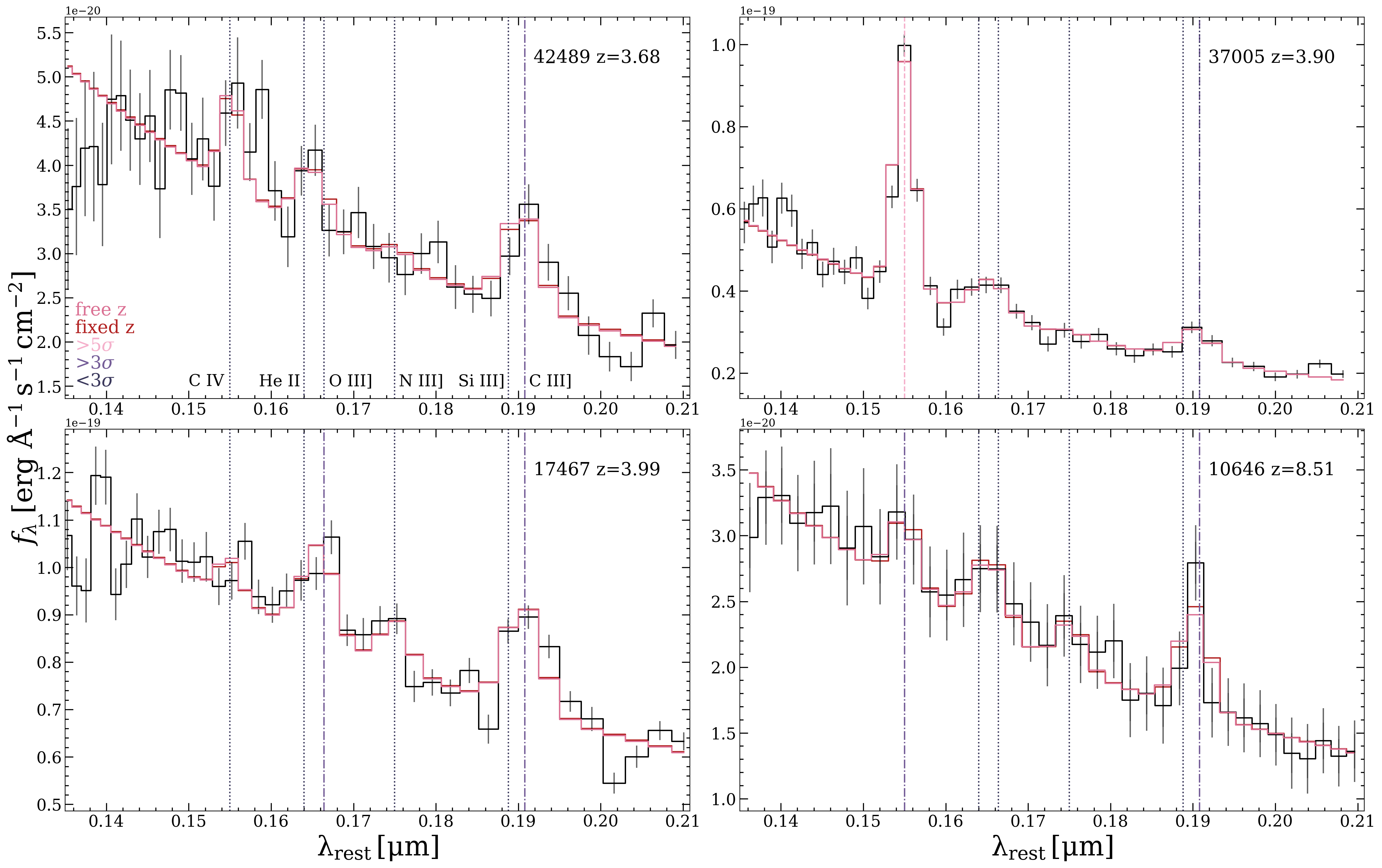}
    \caption{MultiNest fits to C IV through C III] for C III] emitters without He II detections.}
    \label{fig:c3_spectra}
\end{figure*}

\begin{figure*}
    \centering
    \includegraphics[width=\columnwidth]{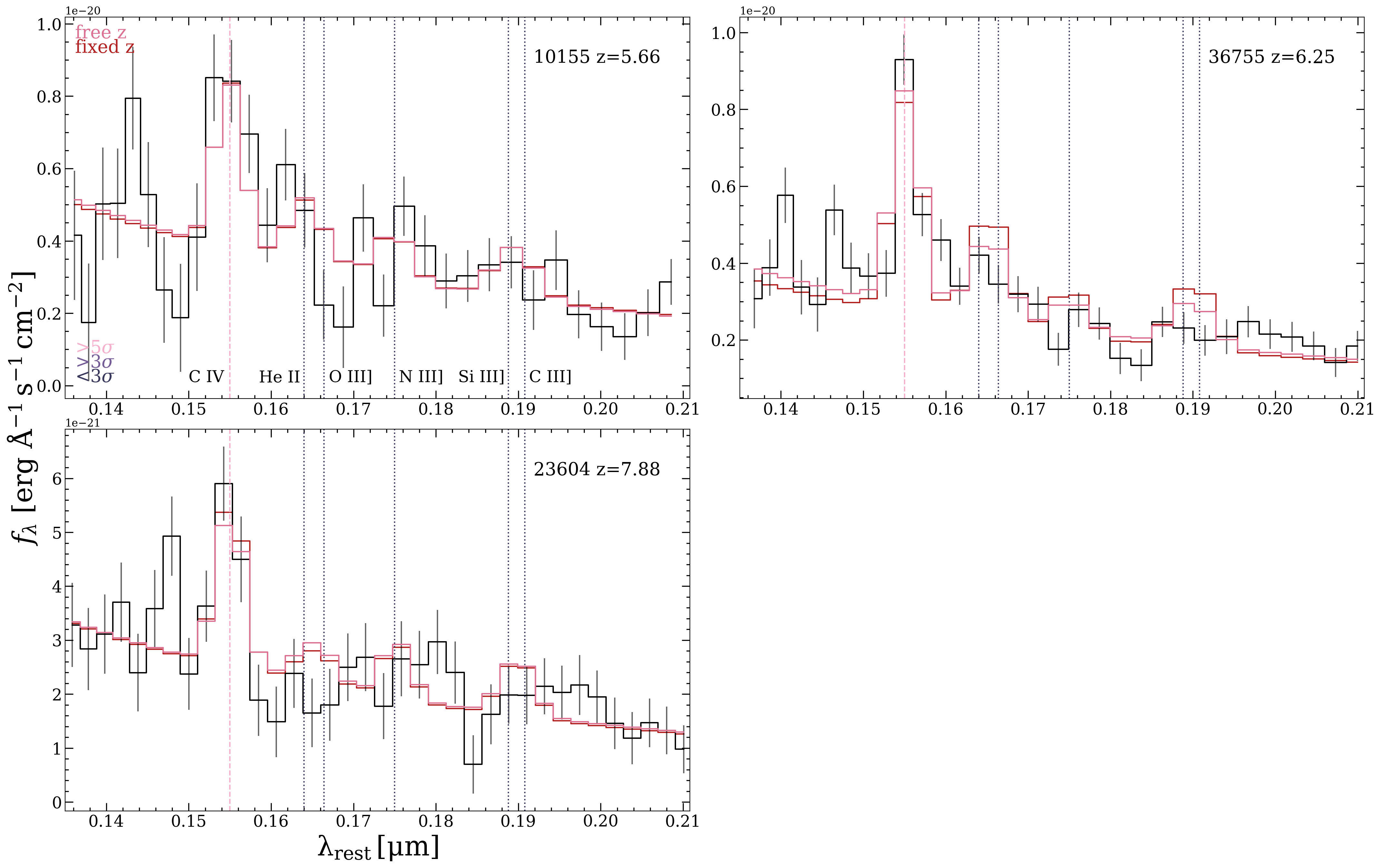}
    \caption{MultiNest fits to C IV through C III] for C IV emitters without He II or C III] detections. 22223, which is noted as a C IV emitter in \cite{Fujimoto2023-22223} and \cite{Topping2024CN}, has a 4$\sigma$ fit, excluding it from the sample. 10155 is also noted as a tentative C IV emitter in \cite{Topping2024CN}.}
    \label{fig:c45_spectra}
\end{figure*}

\end{document}